\documentclass[conference]{IEEEtran}
\usepackage{amsthm}
\usepackage{amssymb}
\usepackage{graphicx}
\usepackage{subfigure}
\usepackage{amsmath}
\usepackage{graphics}
\usepackage{tabularx}
\usepackage[hyphens]{url}
\usepackage{hyperref}
\usepackage{color}

\usepackage{flushend}

\theoremstyle{definition}
\newtheorem{hypothesis}{Hypothesis}

\definecolor{darkblue}{rgb}{0,0,0.7}
\hypersetup{
	colorlinks=true,
	urlcolor=darkblue,
	linkcolor=darkblue,
	citecolor=darkblue
}

\begin{document}
\title{Large-scale Spatiotemporal Characterization of Inconsistencies in the World's Largest Firewall}


\author{
	\IEEEauthorblockN{
		Roya Ensafi\IEEEauthorrefmark{1},
		Philipp Winter\IEEEauthorrefmark{2},
		Abdullah Mueen\IEEEauthorrefmark{1}, and
		Jedidiah R. Crandall\IEEEauthorrefmark{1}
	}
	\IEEEauthorblockA{
		\IEEEauthorrefmark{1} Department of Computer Science \\
		University of New Mexico, USA \\ Email: \{royaen, mueen, crandall\}@cs.unm.edu
	}
	\IEEEauthorblockA{
		\IEEEauthorrefmark{2} Department of Mathematics and Computer Science \\
		Karlstad University, Sweden\\ Email: philwint@kau.se
	}
}

\maketitle
\begin{abstract}
A nation-scale firewall, colloquially referred to as the ``Great
Firewall of China,'' implements many different types of censorship and
content filtering to control China's Internet traffic.  Past work has
shown that the firewall occasionally fails.  In other words, sometimes
clients in China are able to reach blacklisted servers outside of China.
This phenomenon has not yet been characterized because it is infeasible
to find a large and geographically diverse set of clients in China from
which to test connectivity.

In this paper, we overcome this challenge by using hybrid idle scan techniques
that are able to measure connectivity between a remote client and an arbitrary
server, neither of which are under the control of the researcher performing
measurements.  In addition to hybrid idle scans, we present and employ a novel
side channel in the Linux kernel's SYN backlog.  We demonstrate both techniques
by measuring the reachability of the Tor network which is known to be blocked in
China.  Our measurements reveal that \emph{1)} failures in the firewall occur
throughout the entire country without any conspicuous geographical patterns,
\emph{2)} a network block in China appears to have unfiltered access to parts of
the Tor network, and \emph{3)} the filtering seems to be mostly centralized at
the level of Internet exchange points.  Our work also answers many other open
questions about the Great Firewall's architecture and implementation.

\end{abstract}

            
            


\section{Introduction} \label{sec:intro}
More than 600 million Internet users are located behind the world's most
sophisticated and pervasive censorship system: the Great Firewall of China
(GFW)~\cite{china-users}.  Brought to life in 2003, the GFW has a tight grip on
several layers of the TCP/IP model and is known to block or filter IP
addresses~\cite{Winter2012}, TCP ports~\cite{Winter2012}, DNS
requests~\cite{Lowe2007,Anonymous2012,Wright2012}, HTTP
requests~\cite{Clayton2006,Park2010,Crandall2007}, circumvention tools, and even
social networking sites~\cite{Zhu2013}.

\begin{figure}[t]
\centering
\includegraphics[width=0.45\textwidth]{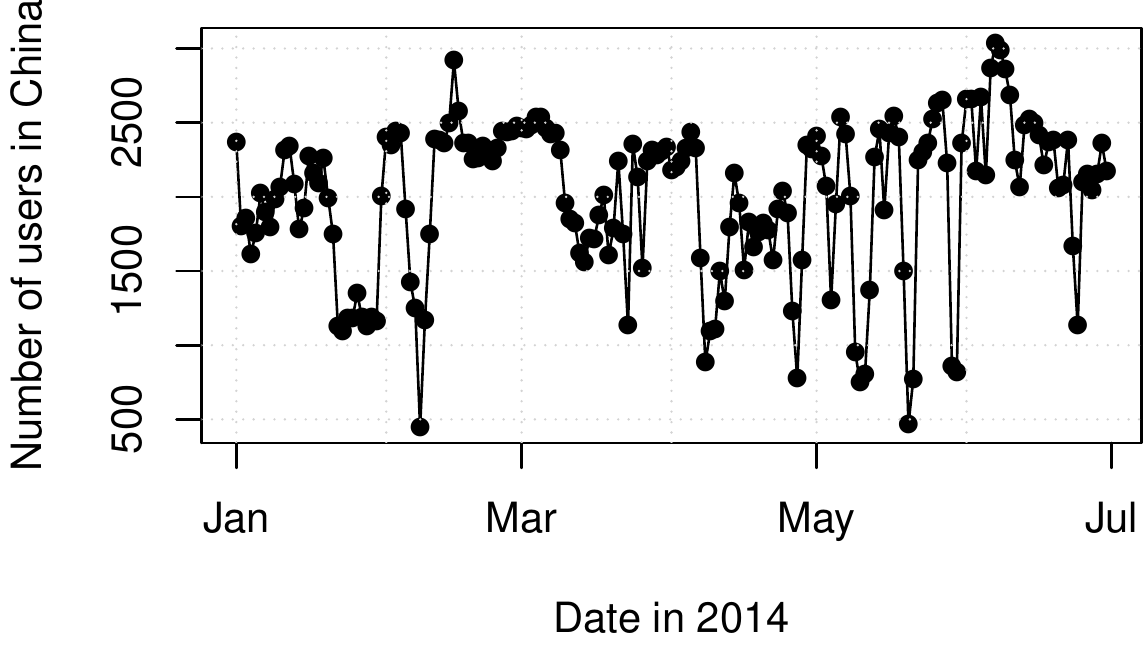}
\caption{The approximate amount of directly connecting Tor users (as opposed to
connecting over bridges) for the first months of 2014.  While the number of
users varies, it rarely exceeds 3,000.}
\label{fig:china_users}
\end{figure}

This pervasive censorship gives rise to numerous circumvention tools seeking to
evade the GFW by exploiting a number of
opportunities~\cite{circumvention-tools}.  Of particular interest is the Tor
anonymity network~\cite{Dingledine2004} whose arms race with the operators of
the GFW now counts several iterations.  Once having had 30,000 users solely
from China, the Tor network now is largely inaccessible from within China's
borders as illustrated in Figure~\ref{fig:china_users}.

The amount of users trying to connect to the Tor network indicates that there
is a strong need for practical and scalable circumvention tools.  Censorship
circumvention, however, builds on \emph{censorship analysis}.  A solid
understanding of censorship systems is necessary in order to design sound and
sustainable circumvention systems.  However, it is difficult to analyze
Internet censorship without controlling either the censored source machine or
its---typically uncensored---communication destination.  This problem is
usually tackled by obtaining access to censored source machines, finding open
proxies, renting virtual systems, or by cooperating with volunteers inside the
censoring country.  In the absence of these possibilities, censorship analysis
has to resort to observing traffic on the server's side and inferring what the
client is seeing.

Our work fills this gap by presenting and evaluating network measurement
techniques which can be used to expose censorship while controlling
\emph{neither the source nor the destination machine}.  This puts our study in
stark contrast to previous work which had to rely on proxies or volunteers,
both of which provide limited coverage of the censor's networks.  By being
mostly independent of source and destination machines, we are able to shed
light on entirely unexplored areas of the Internet.  We evaluate our techniques
by applying them to the Tor anonymity network, thereby handing the Tor Project
practical tools to measure the reachability of their network.  Such tools are
needed because bridges\footnote{Bridges are ``hidden'' Tor relays which are not
listed in the public network consensus.} are frequently blocked in China
without the bridge operators or the Tor Project noticing~\cite{isis-trac}.  Our
work makes it possible to test the reachability of these bridges without having
a vantage point in China.  As a result, the Tor Project is able to learn which
subset of bridges is still reachable and hence undiscovered by the GFW.  This
knowledge facilitates the optimization of bridge distribution~\cite{Wang2013},
\emph{e.g.}, bridges blocked in China are only given out to users outside
China.

Our techniques are currently limited to testing \emph{basic IP connectivity}.
Thus, we can only detect censorship on lower layers of the network stack,
\emph{i.e.}, before a TCP connection is even established.  This kind of
low-level censorship is very important to the censors, however.  For example,
while social media controls on domestic sites in China, such as Weibo, can be
very sophisticated, users would simply use alternatives such as Facebook if the
low-level IP address blocking were not in place to prevent this.  Also, deep
packet inspection (DPI) does not scale as well in terms of raw traffic as does
lower-level filtering.  Nevertheless, we acknowledge that our techniques are
not applicable if censors only make use of DPI to block Tor as it was or is
done by Ethiopia, Kazakhstan, and Syria~\cite{censorwiki}.

We are interested not only in finding patterns in the GFW's \emph{failures}, but
also in gaining a better understanding of how the GFW is \emph{architected}
within China's backbone and provincial networks and whether previously observed
details of its \emph{implementation} are observed throughout the country.  To
this end, we focus our efforts on testing the following hypotheses that will
illuminate the GFW's architecture and implementation.  All hypotheses are with
respect to the filtering of TCP/IP packets based on IP addresses and port
numbers.

\begin{hypothesis}
\label{hypo:synrst}
In general, from any client to any destination if a SYN packet is filtered by
the GFW then a RST with the same source, destination, and port numbers will also
be filtered.  For brevity, we refer to this hypothesis as ``{\it RSTs are
treated the same as SYNs}.''
\end{hypothesis}

\begin{hypothesis}
\label{hypo:allover}
There are no conspicuous geographic patterns in the GFW's failures.  In other
words, failures can occur in any part of the country.  For brevity, we refer to
this hypothesis as ``{\it No geographic patterns in failures}.'' 
\end{hypothesis}

\begin{hypothesis}
\label{hypo:synack}
In general, the GFW blocks Tor relays by dropping SYN/ACK segments with IP
address and port information that matches known Tor relays.  Other types of
filtering seen for Tor relays in China (\emph{e.g.}, dropping SYN segments)
are a negligible fraction of the censorship.  For brevity, we refer to this
hypothesis as ``{\it server-to-client blocking}.''
\end{hypothesis}

\begin{hypothesis}
\label{hypo:persistent}
At least some of the failures of the GFW are persistent, meaning that the
client and server are able to communicate throughout the day.  Note that this
could also be due to intentionally whitelisted destinations, but in this paper
we refer to all cases where clients in China can access Tor relays as
``failures.''  For brevity, we refer to this hypothesis as ``{\it some failures
are persistent}.''
\end{hypothesis}

\begin{hypothesis}
\label{hypo:diurnal}
At least some of the failures of the GFW exhibit diurnal patterns, where a
client and blacklisted server can communicate at some times of the day but not
others.  For brevity, we refer to this hypothesis as ``{\it some failures have
diurnal patterns}.''
\end{hypothesis}

\begin{hypothesis}
\label{hypo:coupleofhops}
In general, packets that are subject to censorship traverse at least one or two
hops, and sometimes more, into China before they are dropped by the GFW.  For
brevity, we refer to this hypothesis as ``{\it blocking is in the backbone}.''
\end{hypothesis}

By testing the above hypotheses, we further increase the public's knowledge
about the GFW and by presenting and evaluating our measurement techniques, we
equip circumvention system developers with a set of tools to analyze and debug
censorship incidents.  In summary, this paper makes the following contributions:

\begin{itemize}
\item We describe the first real-world application of the hybrid idle
scan~\cite{Ensafi2014,extendedversion} to a large-scale Internet measurement
problem, in which we measure the connectivity between the Tor anonymity network
and clients in China over a period of four weeks.
\item We present and evaluate a novel side channel based on the Linux kernel's
SYN backlog which enables indirect detection of packet loss.
\item We increase the community's understanding of how the GFW is architected
and how its blocking of the Tor network looks from different clients all over
China.
\item We publish our code under a free
license to encourage further research.
\end{itemize}

The rest of this paper is structured as follows.  We discuss some background of
the GFW in Section~\ref{sec:backgroundgfw} and our measurement techniques in
Section~\ref{sec:background}, which is then followed by our experimental
methodology in Section~\ref{sec:experimental}.  We analyze the data we gathered
and present results in Section~\ref{sec:results} and proceed with a discussion
of our results in Section~\ref{sec:discussion}.  Related work is covered in
Section~\ref{sec:relatedwork} and the paper is concluded in
Section~\ref{sec:conclusion}.

\section{Motivation and GFW Background} \label{sec:backgroundgfw}

The hypotheses enumerated in Section~\ref{sec:intro} were chosen because we
wanted to address the following open questions about the GFW:

\begin{itemize}
\item Are there geographic or other spatial patterns in the GFW's failures?
This is important because such patterns could be exploited by evasion
technologies if the patterns exist, but if no such patterns exist then evasion
efforts should focus on other aspects of the GFW.
\item Are there temporal patterns in the GFW's failures?  There are many
different evasion efforts that periodically test their methods to see if they
have been detected and blocked by the GFW.  A solid understanding of temporal
patterns (such as diurnal patterns) will help these projects to better
understand the results of their tests.
\item What kinds of packets are filtered in different parts of the country?
This is important, because if an evasion technology is tested in, \emph{e.g.},
Beijing but then fails to work in another part of the country, the developers of
the evasion technology need to understand why.
\item Where in China's Internet backbone does the filtering occur, and what is
the role of routing?  If an evasion technology is being tested from two
different sources in China or two different destinations outside the country,
the developers of the evasion technology may observe two different results for
their tests and they need a good understanding of why this occurs.
\end{itemize}

Now we give more details about what was known before the work presented in this
paper.  A more comprehensive overview of previous work is given in
Section~\ref{sec:relatedwork}.

\subsection{Spatial patterns}

Ensafi \emph{et al.}~\cite{Ensafi2014,extendedversion} found that a small
percentage of tests showed no signs of censorship.  Their tests, like ours, were
taken between clients in China paired with Tor relays outside China.  However,
their experimental methodology was designed to test if the failures in the
censorship observed by Winter and Lindskog~\cite{Winter2012} were also observed
outside of Beijing or not.  Ensafi \emph{et al.} made no attempt to choose
clients or servers so that spatial patterns could be identified.  Our
experiments were specifically designed to identify spatial and geographic
patterns in the GFW's failures.

\subsection{Temporal patterns}

Neither Ensafi \emph{et al.} nor Winter and Lindskog attempted to characterize
temporal patterns in the GFW's failures.  This kind of characterization is
difficult because, for a general understanding of temporal patterns, spatial
patterns must be fully understood.  Otherwise temporal patterns may be specific
to one location.  Also, temporal patterns are difficult to extract from idle
scan measurements because of noise.  This is why, in our experiments, we used
traceroutes from a Tor relay to analyze temporal patterns. 

\subsection{Details of the filtering}

What kinds of packets are filtered?  This is a key question, especially for
evasion technologies that seek to evade the GFW \emph{via} insertion and evasion in the
IP and TCP layers.  Winter and Lindskog described detailed results about
what happens to SYN, SYN/ACK, ACK, and RST packets, but their results were
specific to one location in China: Beijing.  Also, any of their experiments that
required observation on the server were only able to be carried out between
Beijing and one Tor relay in Sweden.  Ensafi \emph{et al.} had more spatial
diversity in their experiments, but because of the nature of their hybrid idle
scan the only packets that can be tested are SYN/ACKs from server to client and
RSTs from client to server.  SYN packets or any kind of stateful connection
cannot be tested with the hybrid idle scan.  All of these limitations in
previous approaches is why our experiments include another---previously
unknown---idle scan that uses the SYN backlog to make more general inferences
with a wider spatial variety.

\subsection{Architecture of the GFW} 

There are generally regarded to be three theories about how the GFW is
architected, posited in technical
papers~\cite{conceptdoppler,Xu2011,Clayton2006,Anonymous2014} or other
media~\cite{atlanticmonthly,wyoming}.  One theory posits that the filtering
occurs at \emph{choke points} where oversea cables carrying international
Internet traffic enter the country.  Another theory is that the majority of the
filtering occurs in \emph{three big Internet exchange points} in Beijing,
Shanghai, and Guangzhou~\cite{bsabouttunnels}, near where international traffic
enters the country but positioned more at central points in China's backbone
network.  A third theory that has been discussed is the possibility that the
filtering occurs---or may increasingly occur as the GFW evolves---at the
\emph{provincial level}~\cite{Xu2011}.

Our results about where the filtering of SYN/ACKs from Tor relays occurs are
largely congruent with Xu \emph{et al.}'s results about where RST injection
based on deep packet inspection occurs.  In their results, CNCGROUP performed
most of its RST injection in the backbone, while CHINANET performed this type
of censorship at the provincial level.  Since their study, CNCGROUP has bought
CHINANET, but the censorship at both the backbone and provincial levels, in
about the same proportions as reported by Xu \emph{et al.}, is also apparent in
our results.  This means that the routers that perform port mirroring for deep
packet inspection are probably the same routers that enforce access controls
such as blocking Tor by source IP address and TCP port.  It also means that
where the filtering occurs has not changed significantly since the study
performed by Xu \emph{et al.}.

In addition to providing more information about where the filtering occurs, our
work presented in this paper raises interesting questions about how the GFW is
architected, both in terms of implementations at routers and in terms of the big
picture.  Winter and Lindskog observed that for Tor relays only the SYN/ACK
from the server is blocked, not the SYN from the client to the server.  One of
our key results in this paper is that this observation also applies to China in
general for a lot of different geographic locations.  This raises a question:
why block SYN/ACKs in the one direction, but not SYNs in the other?

One possible theory might be that the Border Gateway Protocol (BGP) plays a key
role in the censorship by causing all international traffic to flow through the
routers that implement the censorship.  Because the GFW operators are presumably
restricted to announcing BGP routes for autonomous systems (ASes) that are in
China, they can only control routing in the direction of traffic that is
entering China.  Hence SYN/ACKs from Tor relays outside China to clients in
China are blocked almost all the time, while SYNs from clients in China to Tor
relays outside China are much less likely to be blocked.

Another theory is based on speculating about the way the GFW operators monitor
traffic to decide what to block.  In a description of the GFW written in Chinese
by ``Xylon Pan''~\cite{gfwxylonpan}, it is speculated that this is done because
the server in an HTTP connection typically sends a lot more content to the
client than the client sends to the server.  Thus Netflow aggregation in the
server-to-client direction works better, because there is more traffic to be
sampled.  One theory put forward by Xylon Pan is that since the GFW's operators
think about network flows in the server-to-client direction, so they also write
access controls (such as the blocking of Tor by IP address and TCP port) for the
server-to-client direction.

The reason why this one-way blocking property (where SYN/ACKs entering China are
much more likely to be blocked than SYNs leaving China) exists is left for
future work.  The major contribution of our present work in this regard is to
confirm that this property is a general property that is observed all over
China, not just in the one or two locations where previous
tests~\cite{Winter2012,gfwxylonpan}  have been performed.

\section{Networking Background} \label{sec:background}


The research questions we seek to answer require high geographical diversity of
clients in China.  Typically, such a study would only be possible if we could
find and control vantage points in all of China's provinces.  Instead, we
exploit side channels allowing us to detect intentional packet
dropping---without controlling the two affected machines.  In particular, we use
\emph{hybrid idle scans} (see Section~\ref{sec:hybrid_scans}) and \emph{SYN
backlog scans} (see Section~\ref{sec:syn_backlog}).  The idea behind these side
channels as well as their prerequisites are discussed in this section.

\subsection{Side channels in Linux's SYN backlog}
\label{sec:syn_backlog}
A performance optimization in the Linux kernel's SYN backlog can be used to
detect intentional packet dropping.  Half-open TCP connections of network
applications are queued in the kernel's \emph{SYN backlog} whose size defaults
to 256.  These half-open connections then turn into fully established TCP
connections once the server's SYN/ACK was acknowledged by the client. If a
proper response is not received for an entry in the SYN backlog, it will
retransmit the SYN/ACK several times.  However, if the SYN/ACK and its
respective retransmissions are never acknowledged by the client, the half-open
connection is removed from the backlog.  When under heavy load or under attack,
a server's backlog might fill faster than it can be processed.  This causes
attempted TCP connections to not be fully handled while pending TCP connections
time out.  The Linux kernel mitigates this problem by \emph{pruning} an
application's SYN backlog.  If the backlog becomes more than half full, the
kernel begins to reduce the number of SYN/ACK retransmissions for all pending
connections~\cite{LinuxKernelBacklog}.  As a result, half-open connections will
time out earlier which should bring the SYN backlog back into uncritical state.
We show that the Linux kernel's pruning mechanism---by design a \emph{shared
resource}---opens a side channel which can be used to measure intentional
packet drops targeting a server.  This is possible without controlling said
server.

Our key insight is that we can remotely measure the approximate size of a
server's SYN backlog by sending SYN segments and counting the number of
corresponding SYN/ACK retransmissions.  Starting with version number 2.2, the
Linux kernel retransmits unacknowledged SYN/ACK segments five
times~\cite{TCPManpage}.  As a result, we expect to receive the full number of
five retransmissions when querying a service whose SYN backlog is less than
half full.  If, on the other hand, the backlog becomes more than half full, we
will observe less than five retransmissions.  When applied to the problem of
intentional packet dropping, this allows us to infer whether a firewall blocks
TCP connections by dropping the client's SYN or the server's SYN/ACK segment.

It is worth mentioning that a server's backlog state can also be inferred by
coercing it into using \emph{SYN cookies}~\cite{roya}.  A server using SYN
cookies reveals that its SYN backlog is completely full.  However, this
measurement technique is effectively a SYN flood and TCP connections which were
established using SYN cookies suffer from reduced throughput due to the lack of
flow control window scaling.  In contrast to triggering SYN cookies, our technique has no
negative impact on servers or other clients' connections, when applied
carefully.

\subsection{The global IP identifier}
\label{sec:global}
IP identifiers (IPIDs) are unique numbers assigned to IP packets in case they
are fragmented along a path.  The receiving party is able to reassemble the
fragmented packets by looking at their IPID field.  Most modern TCP/IP stacks
increment the IPID field per connection or randomize it, as opposed to
\emph{globally incrementing} it.  A machine with a globally incrementing IPID
keeps a global counter that is incremented by 1 for every packet the machine
sends, regardless of the destination IP address.  Being a \emph{shared
resource}, the IPID can be used by a measurement machine talking to a remote
machine to estimate how many packets the remote machine has sent to other
machines.  Throughout this paper, we refer to machines with globally
incrementing IPIDs as simply machines with ``global IPIDs.''

\subsection{Hybrid idle scan}
\label{sec:hybrid_scans}

\begin{figure*}[t!]
\centering
\includegraphics[scale=0.41]{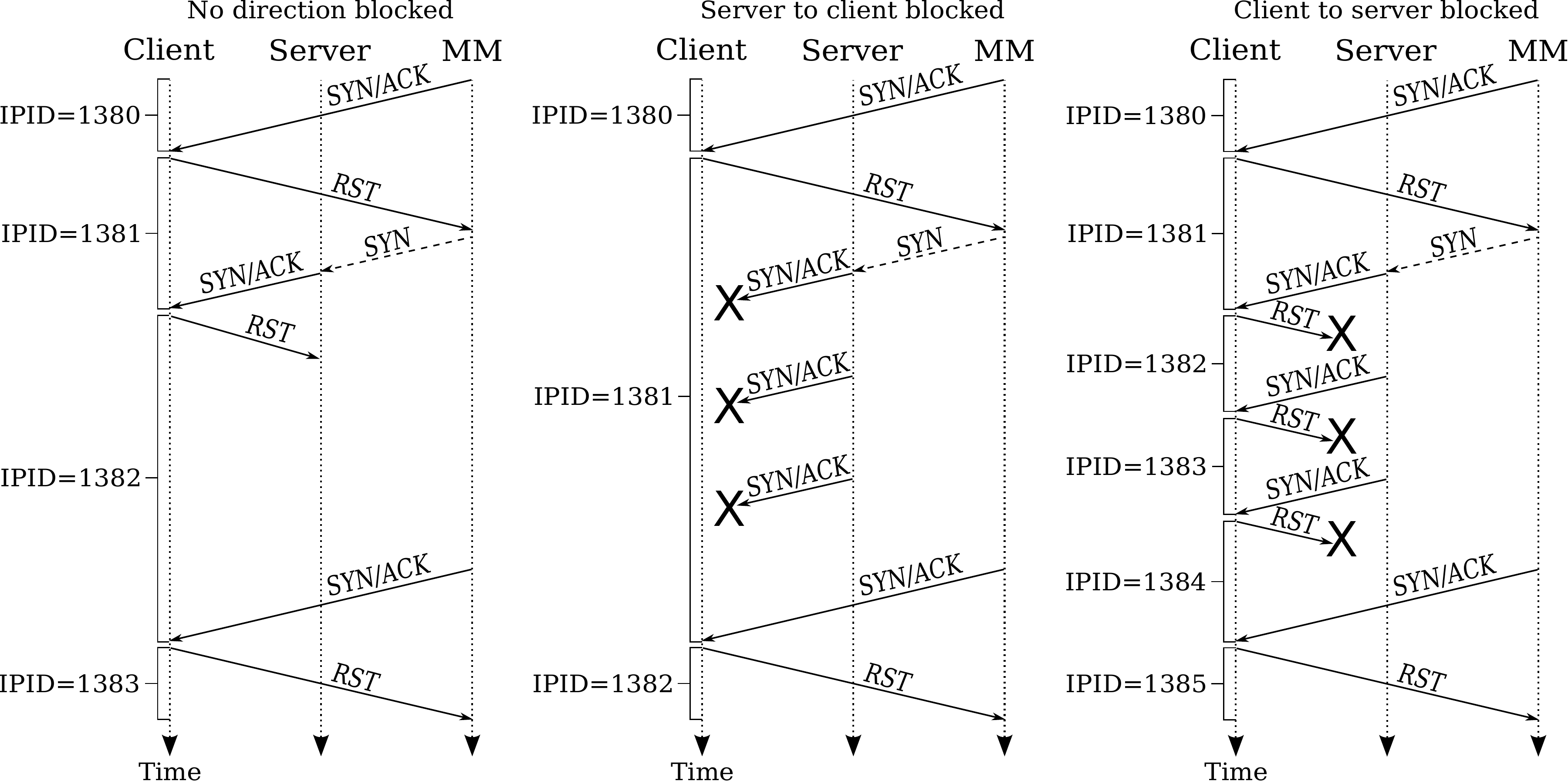}
\caption{Three different cases of packet dropping that our method can detect.
MM is our measurement machine.}
\label{fig:threecases}
\end{figure*}

Ensafi \emph{et al.}~\cite{Ensafi2014,extendedversion} discovered a new method
for remotely detecting intentional packet drops on the Internet via side channel
inferences. Their technique can discover packet drops (\emph{e.g.}, caused
by censorship) between two remote machines, as well as infer in which
direction the packet drops are occurring.  The only major requirements for their
approach are a client with a global IPID and a target server with an open port.
Access to the client or the server is not required.  Conceptually, the hybrid idle scan
technique can turn approximately 1\% of the total IPv4 address
space~\cite{Ensafi2014} into conscripted measurement machines that can be used
as vantage points to measure IP address-based censorship---without having root
access on those machines.  This is why we employ the hybrid idle scan technique
for our geographic study of how Tor is blocked in China.

As shown in Figure~\ref{fig:cases}, the hybrid idle scan implementation queries
the IPID of the client to create a time series.  By sending SYN/ACKs from the
measurement machine and receiving RST responses, the IPID of the client can be
recorded.  The time series is used to compare a base case (when no traffic is
being generated other than noise) to a period of time when the server is sending
SYN/ACKs to the client (because of our forged SYNs).  Recall that the hybrid
idle scan assumes that the client's IPID is global and the server has an open
port.  By comparing two phases, one phase where no SYN packets are sent to the
server and one phase where SYN packets are sent to the server with the return
IP address spoofed to appear to be from the client, the hybrid idle scan
technique can detect \emph{three different cases} (plus an error case), shown
in Figure~\ref{fig:threecases}, with respect to IP packets being dropped by the
network in between the client and the server:

\begin{figure*}[ht]
\centering
\subfigure[Detected as \emph{server-to-client-drop}.]{
	\includegraphics[scale=0.33]{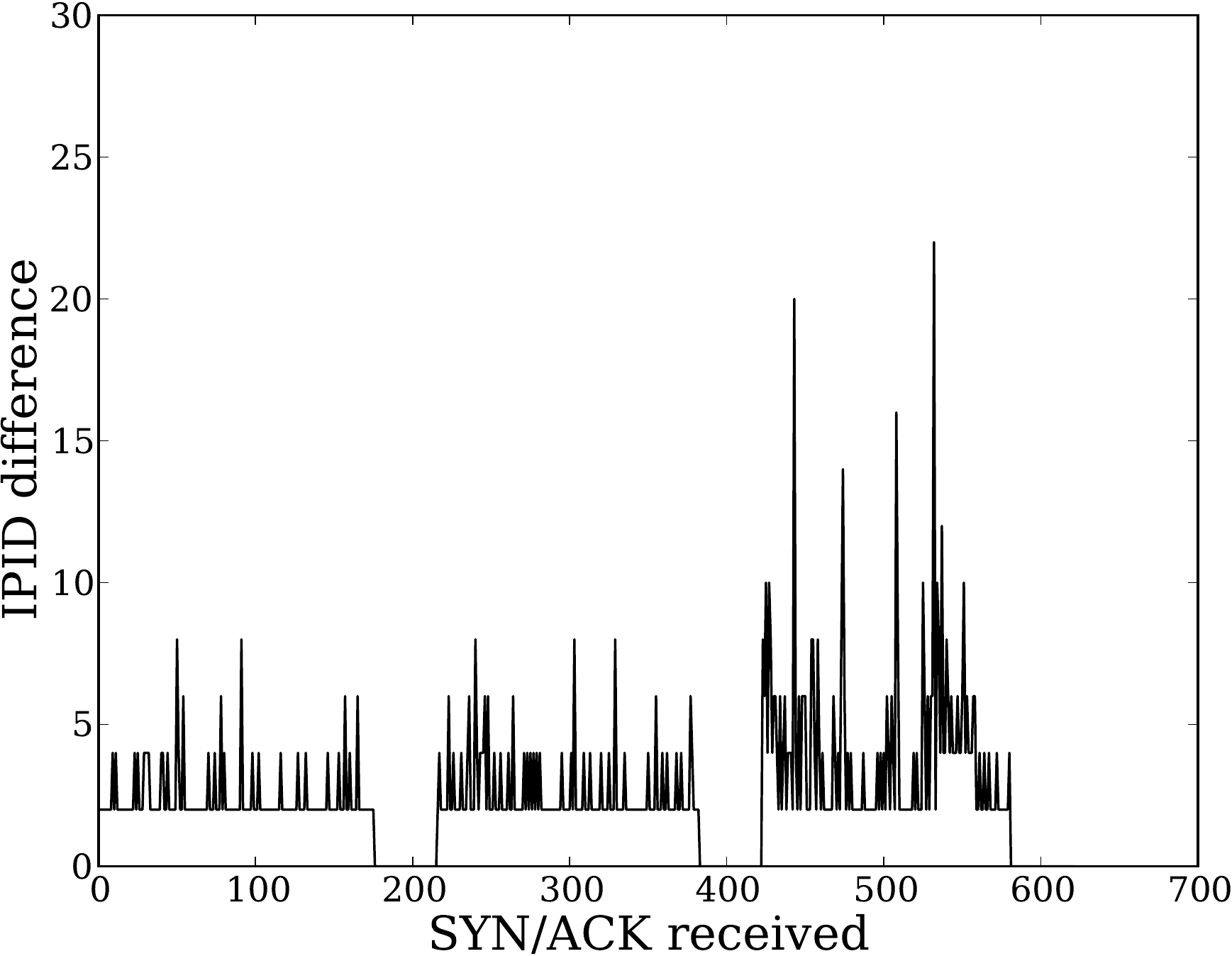}
	\label{fig:case1}
}
\centering
\subfigure[Detected as \emph{no-packets-dropped}.]{
	\includegraphics[scale=0.33]{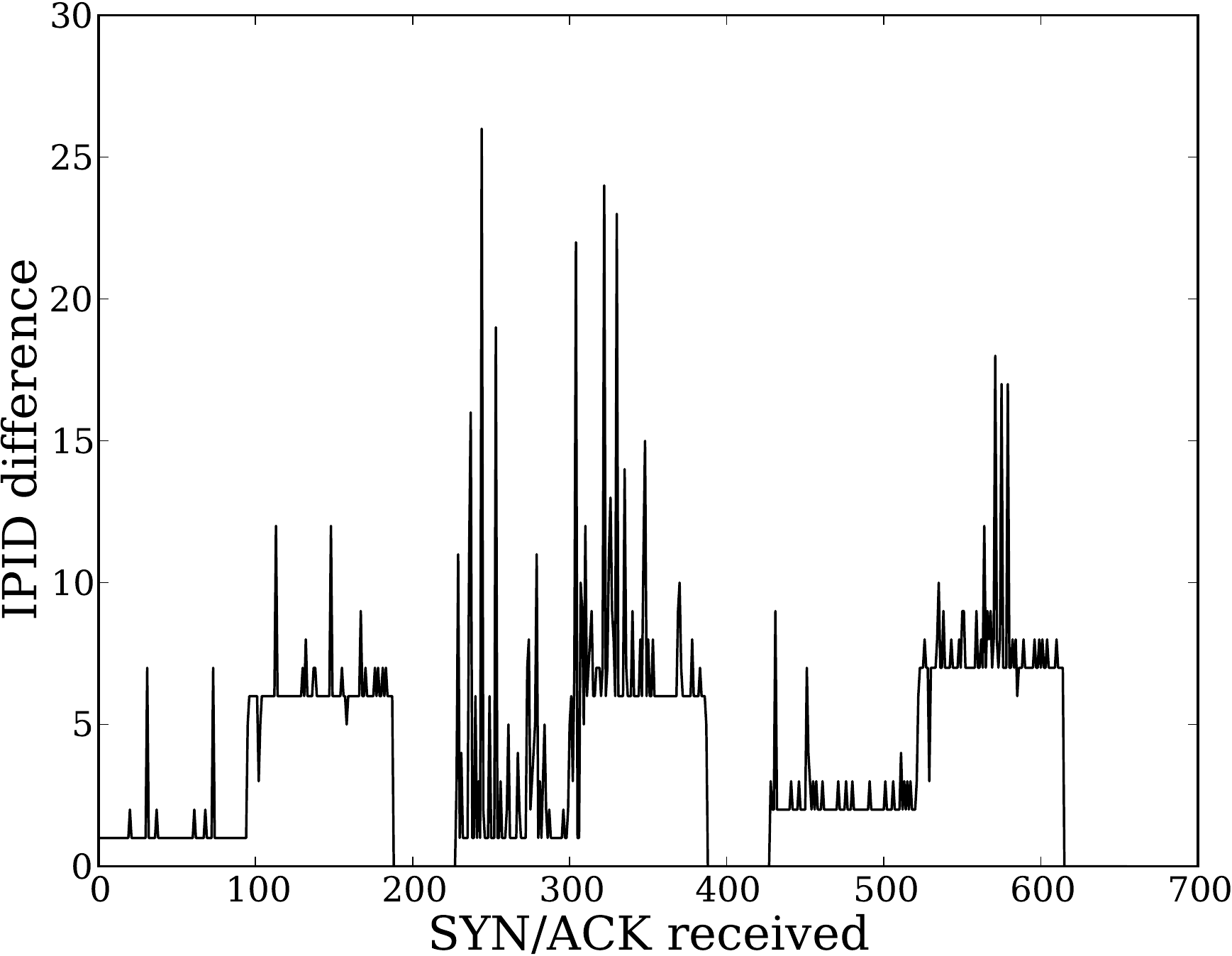}
	\label{fig:case2}
}
\centering
\subfigure[Detected as \emph{client-to-server-drop}.]{
	\includegraphics[scale=0.33]{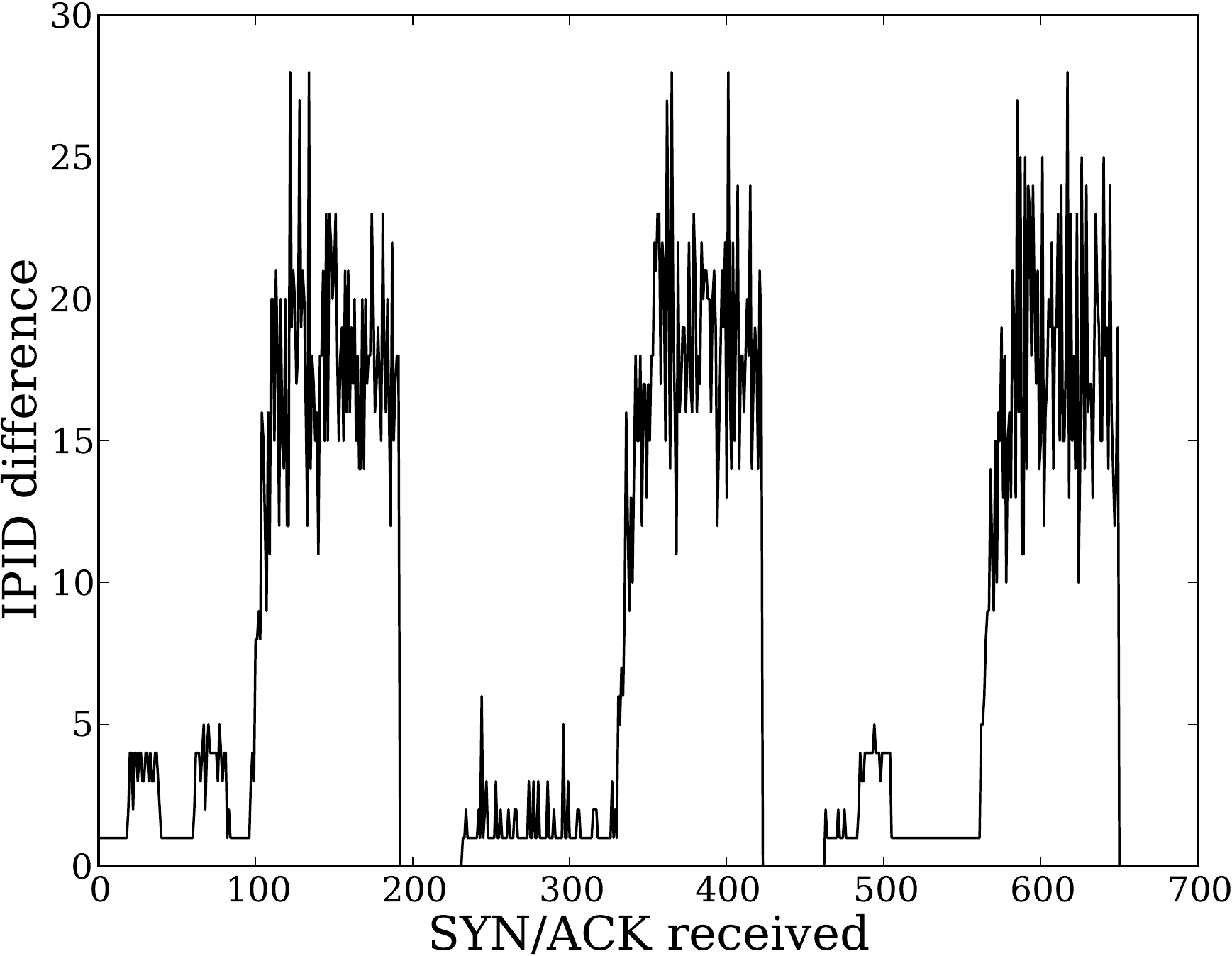}
	\label{fig:case3}
}
\caption{Each subfigure illustrates a time series based on IPID differences for
a specific blocking case.  Despite high amounts of noise, our ARMA modeling can
still detect the blocking case correctly.}
\label{fig:cases}
\end{figure*}

\begin{enumerate}
\item {\bf Server-to-client-dropped:} SYN/ACKs are dropped in transit from the
server to the client causing the client's IPID to not increase at all (except
for noise).  See Figure~\ref{fig:case1}.

\item {\bf No-packets-dropped:} If no intentional packet dropping is happening,
the client's IPID will go up by exactly one.  See Figure~\ref{fig:case2}.  This
happens because the server's SYN/ACK is unsolicited and answered by the client
with a RST segment causing the server to remove the entry from its SYN backlog
and not retransmit the SYN/ACK.

\item {\bf Client-to-server-dropped:} The RST responses sent by the client to
the server are dropped in transit.  In this case, the server will continue to
retransmit SYN/ACKs and the client's IPID will get incremented by the total
number of (re)transmitted SYN/ACKs, which is typically three to six.  See
Figure~\ref{fig:case3}.  This may indicate null routing, the simplest method for
blacklisting an IP address.

\item {\bf Error:} A measurement error happens if networking errors occur during
the experiment, the IPID is found to not be global throughout the experiment, a
model is fit to the data but does not match any of the three non-error cases
above, the data is too noisy and intervention analysis fails because we are not
able to fit a model to the data, and/or other errors.
\end{enumerate}

Auto-regressive moving average (ARMA) models are used to distinguish these
cases.  This overcomes autocorrelated noise in IPID values (\emph{e.g.}, due to
packet loss, packet delay, or other traffic that the client is receiving).
More details about the ARMA modeling are described by Ensafi \emph{et
al.}~\cite{Ensafi2014,extendedversion}.

\subsection{The Tor network}
The Tor network~\cite{Dingledine2004} is an overlay network which provides its
users with anonymity on the Internet.  Tor clients expose a local SOCKS
interface which is used to anonymize TCP streams such as web traffic.  As of
April 2014, the network consists of approximately 4,500 volunteer-run
\emph{relays}, nine \emph{directory authorities}, and one \emph{bridge
authority}.  While the relays anonymize the network traffic of Tor clients, the
authorities' task is to keep track of all relays and to vote on and publish the
\emph{network consensus} which Tor clients need in order to bootstrap.  It is
trivial for censors to download the hourly published network consensus and
block all IP address/TCP port pairs found in it.  Other circumvention systems
suffer from the same problem~\cite{Nobori2014}.

All authorities are hard-coded in the Tor source code and their IP addresses
remain static.  As a result, they constitute attractive choke points for
censors.  In fact, blocking the IP addresses of all nine directory authorities
is sufficient to prevent direct connections to the Tor network.\footnote{Note
that the Tor Project designed and implemented so-called bridges to
tackle this very problem but the details are outside the scope of this work.}
Our study focuses on the reachability of the authorities and relays, as it is
known that the GFW is blocking them~\cite{Winter2012}.  Our focus is on
gathering more details about this blocking and characterizing it with a
large-scale spatiotemporal study.

\section{Experimental Methodology} \label{sec:experimental}

In this section, we describe the challenges our experimental methodology was
designed to address, the data sets we collected, how our measurements help us to
test the hypotheses enumerated in Section~\ref{sec:intro}, and other issues.

\subsection{Encountered challenges}
\label{sec:challenges}

Over the course of running our experiments and analyzing our data, we faced a
number of challenges which we discuss here.

\textbf{Churn in the Tor network}: While the size of the Tor network does not
vary considerably over a short period of time, the network's \emph{churn rate}
can render longitudinal studies difficult.  For example, the median size of
Tor's network consensus (\emph{i.e.}, the number of Tor relays in the network)
in March 2014 was 5,286.  In total, however, March has seen 13,343 \emph{unique
relays}---many of which were online for only hours.  To minimize the chance of
selecting unstable Tor relays for longitudinal studies, only relays having
earned the ``Stable'' flag should be considered~\cite{dirspec}.  Furthermore,
the relay descriptor archives could be examined to calculate a relay's
reachability over time~\cite{deschist}.  We selected only Tor relays that had an
uptime of at least five days, and filtered out all data points where a node
appeared to have left the network.  After having run our experiments, we
removed one Tor relay in Argentina from our data because its Tor and web ports
switched during our experiments.

\textbf{Geolocation of routers}: For geolocating routers, we used MaxMind's
GeoIP2 City database~\cite{geoip2city}.  As of April 2014, this database lacks
accurate geolocation information for backbone routers in China.  While
provincial routers can typically be mapped to their province based on whois
records, backbone routers are all mapped to the same bogus location at latitude
35 and longitude 105 which resides in an unpopulated area in central China.  We
also used MaxMind for geolocating clients, for which it is fairly accurate.
For the location of routers, we used a combination of \emph{whois information}
and \emph{round-trip delays} per hop.  We discarded hops in our data that have
whois records from China but are actually in Hong Kong or Pasadena, CA (where
ChinaNet has a Point of Presence).

\textbf{Diurnal patterns}:  For most measurements in this paper, we measured
once per hour throughout the day.  This avoids bias and distortion.  For
example, if we measured one set of clients in the morning and one set at night,
differences between the two sets of clients may be due to different traffic
patterns at the different times of day and not a property of the different set
of clients.  Thus we always randomize the order of our experiments when possible
and repeat all measurements every hour for at least one full day.

\subsection{Experimental design and setup}

Over the course of our experiments, we made use of three sets of Linux-based
measurement machines in the U.S., China, and Europe.  These three sets of
machines correspond to the three main datasets that we collected.

\textbf{Machines in the U.S.:} The three machines used for our hybrid idle scans
(see Section~\ref{sec:hybrid_scans}) and SYN backlog scans (see
Section~\ref{sec:syn_backlog}) were located
at our university campus (UC) at the University of New Mexico.  All machines
had a direct link to a research network which is free from packet filtering and
does not conduct egress filtering to block spoofed return IP addresses.
Furthermore, the UC measurement machines have IP addresses that are not bound
to any interfaces in order to eliminate unsolicited network packets.  For
example, a measurement machine's kernel should never send a RST when it
receives a SYN/ACK.  The data set collected using the hybrid idle scan from
these machines is a large-scale geographic pairing of many clients (in China
and other countries) with many Tor relays and web servers around the world
(mostly outside China).  It complements the other data sets discussed below
because it gives a complete cross-section of censorship between many clients
and many servers.  This data will be used to test Hypotheses~\ref{hypo:allover}
({\it no geographical patterns in failures}) and~\ref{hypo:persistent} ({\it
some failures are persistent}). 

\textbf{VPS in China:} We rented a virtual private system (VPS) in China.  The
system was located in Beijing (AS 23028) and was used for our SYN backlog scans
discussed in Section~\ref{sec:syn_backlog}.  Our VPS provider employed a
transparent and stateful TCP proxy in front of our VPS which silently dropped
unsolicited segments.  We carefully implemented our SYN backlog scans so they
first established state whenever necessary to be unaffected by the TCP proxy.
These SYN backlog scans provide a dataset that speaks to our assumptions about
how China blocks Tor.  It complements the hybrid idle scan data set because,
although the measurements are from a single client in China, it allows us see
exactly how that client experiences the censorship.  This data will be used to
test Hypotheses~\ref{hypo:synrst} ({\it RSTs are treated the same as SYNs})
and~\ref{hypo:synack} ({\it server to client blocking}).

\textbf{Tor relay in Europe:} We used a long-established Tor relay
at Karlstad University in Sweden for our traceroute measurements discussed in
Section~\ref{sec:traceroutes}.  The relay has been part of the Tor network for
several months, and using our VPS we manually verified it to be blocked in
China.  This data set shows blocking between one Tor relay and many clients in
China.  It complements the hybrid idle scan data set because access to the Tor
relay allows us to collect more details about the blocking.   This data will be
used to test Hypotheses~\ref{hypo:persistent} ({\it some failures are
persistent}), \ref{hypo:diurnal} ({\it some failures have diurnal patterns}),
and~\ref{hypo:coupleofhops} ({\it blocking is in the backbone}).

We now present our probing infrastructure as well as our measurement methodology
used to investigate the theories posited in Section~\ref{sec:background}.

\subsubsection{Hybrid idle scans}\label{sec:exp_hybrid}

Recall that by using hybrid idle scans, we have more freedom in choosing clients
in different regions to test their reachability to different servers.  Our goal
is to determine blocking of Tor relays (outside of China) from the perspective
of a large and geographically diverse set of clients (within China).

We are interested in knowing whether there exist different experiences of the
censorship of Tor for different users in different regions.  Past work showed
that a small fraction of all Tor relays was accessible from a single vantage
point in Beijing~\cite{Winter2012}, but what about the rest of the country?  Key
questions are: how does the GFW's architecture and China's routing affect
censorship in different regions?

\textbf{IP address selection}: We selected clients in China (CN), North America
(NA), and Europe (EU).  In order to be able to select random IP addresses in
China without favoring specific locations---especially large cities featuring a
vast number of allocated IP addresses---we divided the map of China into $33 *
65$ cells corresponding to one degree of latitude and longitude.  We filled this
grid with all IP addresses in MaxMind's database that were documented to be in
China.  Then, we collected IP addresses by randomly selecting a cell from our
grid after checking that they employed global IPIDs.  In an analogous manner,
clients from the EU and NA were chosen by horizontally scanning these regions.
After 24 hours, we gathered a pool of IP addresses that belonged to machines
with a global IPID.  Then, we continually checked the selected IP addresses for
a 24-hour period to discard IP addresses that changed global IPID behavior, went
down, or were too noisy.  At the end we had 11 NA, 7 EU, and 161 CN
clients to use for our measurements. 

Servers were chosen from three groups: Tor relays, Tor directory authorities,
and web servers.  Tor relays were downloaded from a Tor relay status
list~\cite{torstatus}.  We only selected relays with an uptime greater
than five days.  In order to select Tor relays in geographically diverse regions,
we selected 10 Tor relays from Europe, 13 from the United States, 20 from
Russia, and 101 from other countries.  This way, our selected Tor relays were
not biased toward Europe or the U.S., which exhibit more relays per capita than
other regions.  The 10 Tor authorities were obtained from the Tor source code.
Web servers were chosen randomly from Alexa's top 50 websites in
China~\cite{alexaCN}.  All web server and Tor relay IP addresses were checked
hourly to make sure that they stayed up for at least 24 hours before being
selected for our measurement.

\begin{figure*}[t]
\centering
\includegraphics[width=0.7\textwidth]{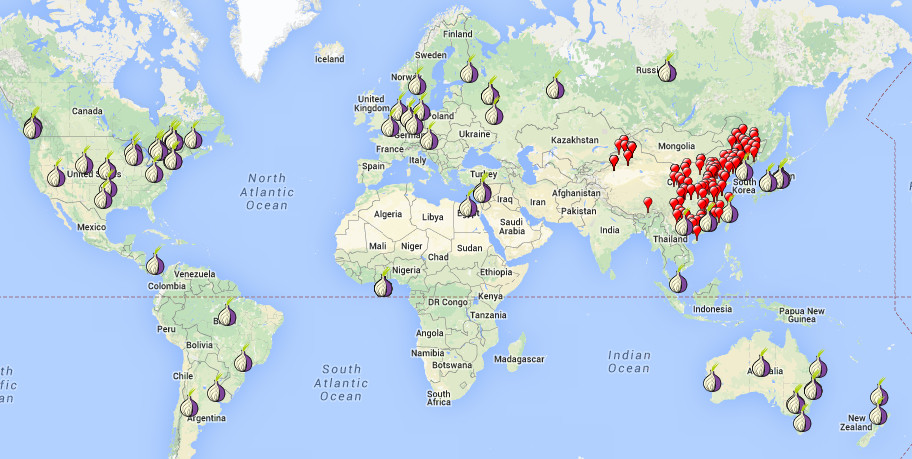}
\caption{The geographic distribution of all tested Tor relays (shown as
onions) and of our global IPID clients in China (shown as red marks).  Note that
outside of Xinjiang the west of China has very little Internet penetration,
which is why we have few data points in this region and the distribution is
biased towards the eastern parts of China.
(Map data {\copyright} 2014 Google, INEGI)}
\label{fig:CSmap}
\end{figure*}

The geographic distribution of our Tor relays as well as all clients in China is
illustrated in Figure~\ref{fig:CSmap}.

\textbf{Creating a complete bipartite graph}: We used three machines at UC (our
university campus) to run the hybrid idle scan experiments.  We started the
experiments with 180 clients and 176 servers.  Each day 20 clients and
approximately 20 servers were selected for each of the machines.  For 22
hours\footnote{Two hours per day were reserved for server data
synchronization.}, every hour, we performed the hybrid idle scan for each
possible pair of client and server.  Every ``scan round'' performs: \emph{1)}
two minutes of hybrid idle scans, \emph{2)} 30 seconds of sending RSTs to clear
the server's backlog, and \emph{3)} five seconds of testing the client to
assure that they remained online and kept their global IPID.  Similar checks
are performed to ensure that servers remain online throughout each experiment.
At any given time, each IP address (client or server) was involved in only one
test.  After 27 days, each client's reachability was tested to all servers,
\emph{i.e}, \emph{our clients and servers created a bipartite graph}.  For more
details about the experiment design refer to Ensafi~\emph{et
al.}~\cite{Ensafi2014,extendedversion}.

\textbf{Pruning the data:} We used the selected IP addresses throughout our
experiments.  Naturally, some of the hosts went down or were occasionally too
noisy.  Also, the host behind an IP address can change, \emph{e.g.}, a client
with a global IPID might lose its DHCP lease and get replaced with a client
running a random IPID.  To account for these issues, we perform tests
throughout our experiments which cull out data points
where basic assumptions are not met.  For every server involved in the
experiment, we had two checks: liveliness and the stable Tor flag test.  After
each scan, for five seconds we sent five SYN segments per second using UC's
unbound IP address.  The data point passed the liveliness test only if it
retransmits three or more SYN/ACKs.  Also, if the server was a Tor relay, we
verified that the relay was assigned the ``Stable'' flag (cf.
Section~\ref{sec:challenges}).

For every client, for five seconds, we sent five SYN/ACKs per second using UC's
unbound IP address.  We expect the client to respond with RST segments totaling
in number to more than half the number of sent SYN/ACKs.  If this is the case
then the data point passes the client's liveliness test.  The results of a scan
were allowed into the data set only if both the client and server passed their
checks.  Note that each data point is one client and one server tested one time
in a given hour.  There was a several-hour network outage that caused a hole in
a portion of one day of our data.

After culling out data that did not meet our basic assumptions, we were left
with $36\%$ of the total data collected.  This $36\%$ is the data described in
Section~\ref{sec:results} and used for our analysis.

\subsubsection{Backlog scans}
\label{sec:backlog_scans}
\begin{figure*}[ht]
\centering
\subfigure[SYN scan to infer whether SYN segments from VPS reach Tor. ``MM'' is
our measurement machine.]{
	\includegraphics[scale=.55]{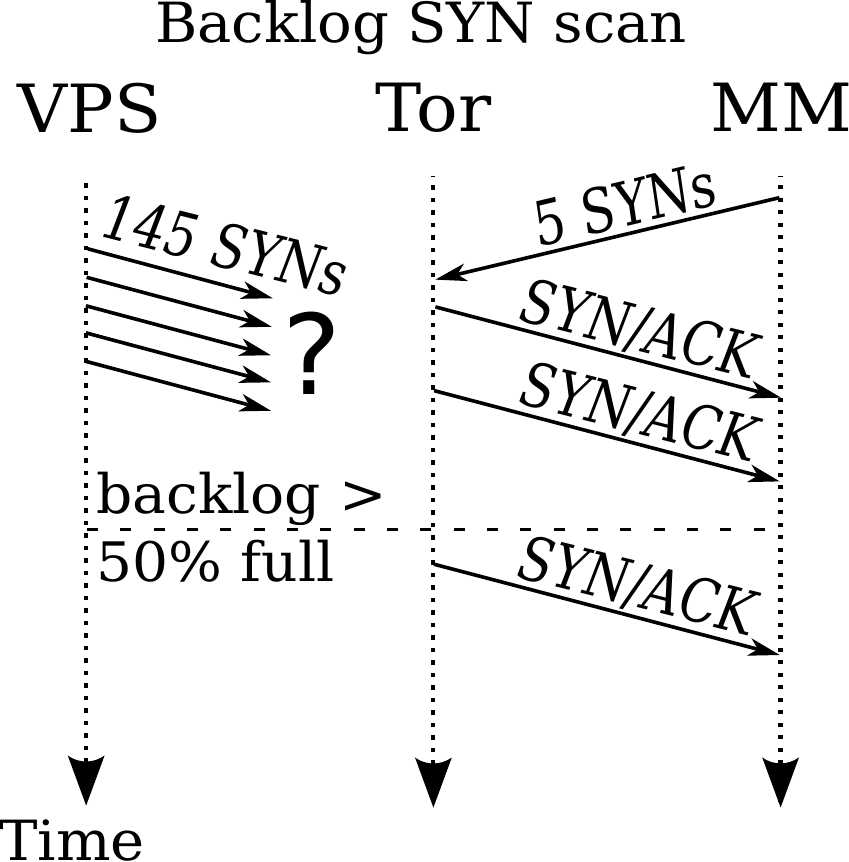}
	\label{fig:syn_backlog_scan}
}
\hspace{1in}
\subfigure[RST scan to infer whether RST segments from VPS reach Tor.  ``MM''
is our measurement machine.]{
	\includegraphics[scale=.55]{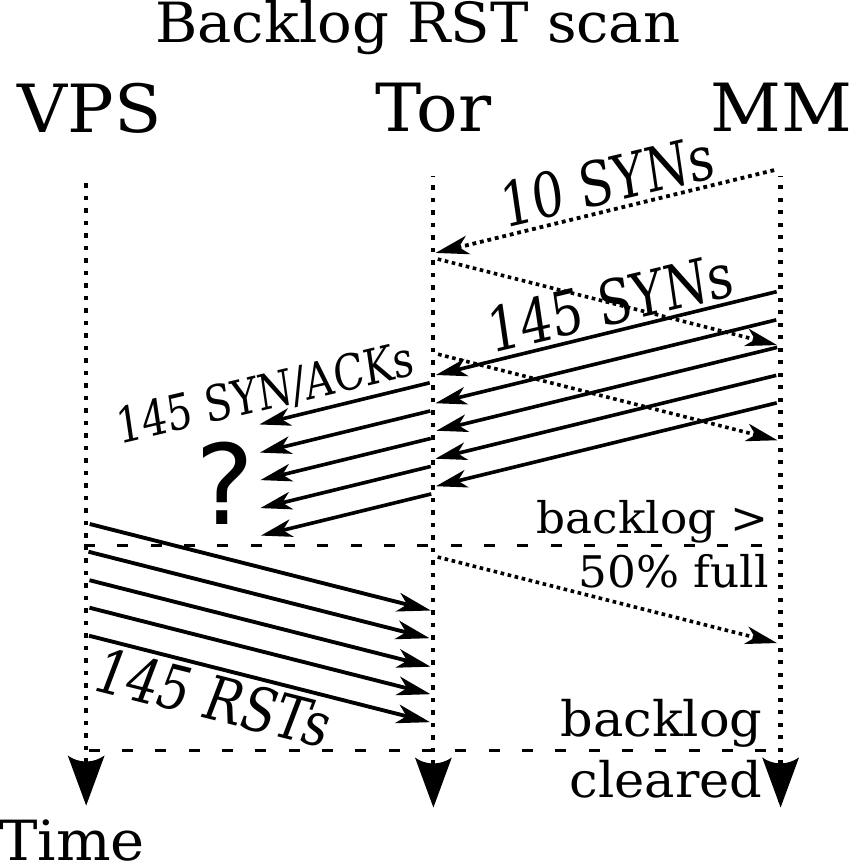}
	\label{fig:rst_backlog_scan}
}
\caption{The two types of backlog scans we employ.  The purpose of these scans
	is to verify if \emph{1)} SYN segments from China reach a Tor relay and if
	\emph{2)} RST segments from China reach a Tor relay.}
\label{fig:backlog_scans}
\end{figure*}

After having presented the underlying side channel in
Section~\ref{sec:syn_backlog}, we now discuss the implementation of our two
backlog scan types which can answer two questions, \emph{1)} ``Do SYN segments
from China reach a Tor relay?'' and \emph{2)} ``Do RST segments from China
reach a Tor relay?''.  Basically, we answer both questions by first
transmitting crafted TCP segments to a relay, thus manipulating its SYN
backlog, and then querying its backlog size by counting the relay's SYN/ACK
retransmissions.  The conceptual implementation of both scan types is
illustrated in Figure~\ref{fig:backlog_scans}.

\textbf{SYN scan}: The SYN scan---depicted in
Figure~\ref{fig:syn_backlog_scan}---is started by MM by sending five SYN segments
to Tor in order to infer the relay's backlog size when under
stress.\footnote{We transmit five SYN segments rather than just one to account
for packet loss.}  After a delay of approximately 500 ms, VPS proceeds by
sending 145 SYN segments whose purpose is to fill the relay's backlog by more
than half. Recall that the backlog size defaults to 256, so we only fill the
backlog to 59\%.  That way, we can make the Tor relay's kernel prune MM's SYN
segments, thus reducing their retransmissions.  Finally, MM knows that VPS's
SYNs reached the relay if the number of SYN/ACK retransmissions for its five SYNs
is lower than five.  Otherwise, VPS's SYNs did not reach the relay.  This type
of inference is necessary because, most of the time, China's GFW drops SYN/ACKs
from known Tor relays.

\textbf{RST scan}: Our RST scan incorporates an additional step but is based on
the same principle.  As illustrated in Figure~\ref{fig:rst_backlog_scan}, MM
starts by sending 10 SYN segments whose purpose is, analogous to the SYN scan,
to monitor the relay's backlog size.  Afterwards, MM proceeds by sending 145
spoofed SYN segments with VPS's source address.  Note that we cannot send the
SYN segments from VPS as they might be blocked.  By sending spoofed SYN
segments from an unfiltered network link, we can ensure that the segments reach
the Tor relay.  Upon receiving the SYN segment burst, the relay replies with
SYN/ACK segments which we expect to be dropped by the GFW.  In the final step,
VPS sends a burst of RST segments to the Tor relay.  The RST segments are
crafted so that every RST segment corresponds to one of the relay's
SYN/ACK segments.  The purpose of the RST burst is to terminate all half-open
connections, thus clearing the relay's backlog.  Based on how many
retransmissions we observe for the 10 ``probing SYNs'', we can infer whether
the RST segments were dropped by the GFW or not.  Receiving five retransmissions
means that the backlog was not cleared and the RST segments were dropped.
Receiving less than five retransmissions means that the backlog was successfully
cleared and the RST segments were not dropped by the GFW.  This kind of
inference is necessary because machines outside China cannot measure directly
what happens to RST packets sent from China, and machines inside China are very
limited in their ability to infer what is happening on blocked IP address/TCP
port pairs.

\textbf{Implementation:} We implemented our scans using a collection of bash
scripts and a patched version of the tool hping3~\cite{hping}. Accurate timing
was crucial for our experiments. To keep the clock of our machines
synchronized, we used the tool ntp which implements the network time protocol.
Recall that the SYN backlog behavior we are exploiting is limited to Linux
kernels (cf. Section~\ref{sec:syn_backlog}).  As a result, our scans targeted
the subset of 94 out of our 144 Tor relays which are known to run Linux.  Tor
relays periodically publish their server descriptors---which includes their
operating system---to all directory authorities so there is no need for us to
guess the operating system of Tor relays.

\textbf{Pruning the data:} By pruning the backlog scan data, we aim to make
sure that the relay runs an unmodified Linux TCP/IP stack.  After scanning a
relay, we send three ``baseline SYNs'' to it in order to query its original
amount of SYN/ACK retransmissions.  First, we discard scans in which the relay
never sent five SYN/ACK retransmissions, Linux's default value since version
2.2.  For example, we found embedded Linux relays which always retransmit
SYN/ACK segments four times, regardless of their backlog size.  Second, we also
discard scans whose SYN/ACK retransmissions do not exhibit Linux's exponential
backoff behavior.  Third and finally, we discard scans where the relay was
offline or other networking problems occurred.  These three pruning steps
discarded 774 out of all 2,094 scans (37\%).

\subsubsection{Traceroutes into China}
\label{sec:traceroutes}
We want to learn if there are \emph{unfiltered routes} leading into China.  To
investigate this question, we used our Tor relay
in Europe to run
traceroutes to numerous destinations in China.  After a country-wide scan, we
obtained a list of 3,934 IP addresses in China that responded to SYN/ACKs and
were distributed geographically in a diverse way, which served as our traceroute
destinations.  For every IP address, we ran two TCP traceroutes; one whose TCP
source port was equal to the filtered Tor port 9001 and one whose TCP port was
set to the unused and unfiltered port 9002.  The traceroutes had both their SYN
and ACK bit set.  We used a slightly modified version of the tool
hping3~\cite{hping} to run the traceroutes as it allowed us to send TCP segments
with a source port which is bound by the Tor process.\footnote{We modified the
tool to constantly increase the TTL of outgoing TCP segments.  The default
behavior is to wait for every hop to reply with a ``TTL exceeded'' ICMP
message.} Starting on 4 May 2014, we ran the traceroutes on an hourly basis
for two days, resulting in a total of $3,934 \cdot 24 \cdot 2 \cdot 2 = 377,664$
traceroutes.  We determined where the traceroutes entered China using whois and
round-trip time information.  We culled out a small amount of data that did not
enter China through a known backbone network, since all such data either
appeared to enter China in Pasadena, California (a case we can handle but will
require deeper analysis into whois records) or was destined for clients that we
determined to actually be in Hong Kong.

\subsection{Good Internet citizenship}
We took several steps to devise our scans to be minimally invasive.  First, we
set up a web server on our measurement machines whose index page informed
visitors about our experiments.  The page contained our contact information to
provide alarmed network operators with an opportunity to contact us and opt out
of our measurements.  Furthermore, we carefully designed our measurements so
that it is very unlikely that they harmed any computers or networks.
Throughout the lifetime of our experiments, we did not receive any complaints.
We discuss ethical aspects of our measurements in Section~\ref{sec:ethics}.

\section{Analysis and Results}\label{sec:results}
We now analyze the three data sets we gathered; the hybrid idle scans, the
backlog scans, as well as the traceroutes into China.

\subsection{Hybrid idle scans}

The hybrid idle scan data was collected from 15 March 2014 to 10 April 2014.
One client was removed from the data because we determined that it was in Hong
Kong and as a result not subject to the GFW's filtering.
 
Table~\ref{tab:results} shows the results of our hybrid idle scans. The
column $S \rightarrow C$ is short for {\bf Server-to-client-dropped}, {\it None}
means {\bf No-packets-dropped}, $C \rightarrow S$ means {\bf
Client-to-server-dropped}, and {\it Error} simply means {\bf Error}.  In the
table's rows, {\it CN} is short for China, {\it EU} means Europe, and {\it NA}
means North America.  As for the server types, {\it Tor$-$Dir} is a Tor
directory authority, {\it Tor$-$Relay} is a Tor relay, and {\it Web} is a web
server.  Our results confirm that, in general, SYN/ACKs entering China from
blacklisted IP address/TCP port pairs are blocked.  Some web servers were
censored, and some Tor nodes were censored outside China.  This is to be
expected because even in countries that do not perform nation-scale Internet
censorship, organizations frequently take steps to filter material such as
pornography or file sharing sites.  Note that highly popular websites often
contain material that is subject to censorship.

\begin{figure}[t]
\centering
\includegraphics[width=0.47\textwidth]{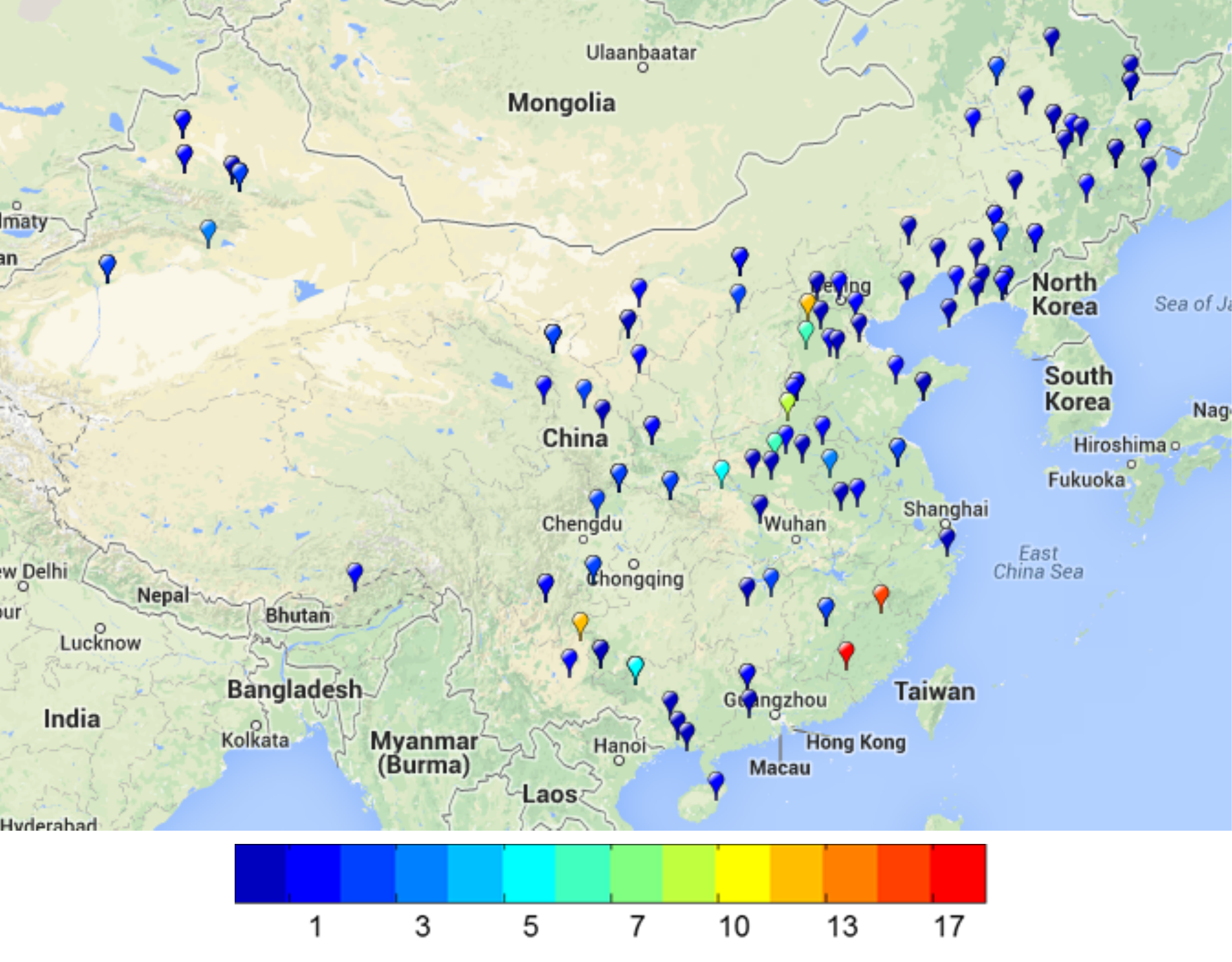}
\caption{The color temperature for clients corresponds to the number of
observed {\bf No-packets-dropped} cases over the entire experiment.  No
geographic or topological pattern is visible.  Instead, the distribution
matches the geographic Internet penetration patterns of China.
(Map data {\copyright} 2014 Basarsoft, Google, ORION-ME, SK planet, ZENRIN)}
\label{fig:Case2Heat}
\end{figure}

The most interesting result from the hybrid idle scans is that the {\bf
No-packets-dropped} case was measured all over the country without any
noticeable geographic pattern.  The geographic distribution of observed {\bf
No-packets-dropped} cases is shown in Figure~\ref{fig:Case2Heat}. The case
distribution closely matches the distribution of our clients which, in turn,
matches the geographic Internet penetration patterns of China.  This means that
the failures in China's IP address/TCP port blacklisting mechanisms are not
limited to one region or one network block.  We provide a more thorough
analysis in Section~\ref{sec:mueen}, which confirms
Hypothesis~\ref{hypo:allover}.

We also observed that in many cases these filtering failures are
persistent and \emph{last throughout the day}.  We witnessed four client/server
pairs where all 22 measurements in a day returned {\bf No-packets-dropped}. We
redacted the clients' 16 least significant bits:


\noindent
Client 58.193.0.0 (CN) $\rightarrow$ server 198.96.155.3 (CA) \\
Client 58.193.0.0 (CN) $\rightarrow$ server 161.53.116.37 (HR) \\
Client 58.193.0.0 (CN) $\rightarrow$ server 128.173.89.245 (US) \\
Client 121.194.0.0 (CN) $\rightarrow$ server 198.96.155.3 (CA) \\

This would give evidence towards Hypothesis~\ref{hypo:persistent}, but our
traceroute results reveal that CERNET does not perform the type of blocking we
are measuring at all so later in this section we will discuss similar failures
in commercial networks.  Clients 58.193.0.0 and 121.194.0.0 are part of the
Chinese Educational and Research Network (CERNET).  Server 198.96.155.3 is a
long-established Tor exit relay at the University of Waterloo.  161.53.116.37
and 128.173.89.245 are Tor relays in Croatia and the U.S., respectively.  There
were also many instances where client/server pairs showed {\bf
Server-to-client-dropped} for most of the day but also showed {\bf
No-packets-dropped} once or a handful of times.

\begin{table*}[ht]
\small
\caption{Results from the hybrid idle scan measurement study.\label{tab:results}}
\centering
\begin{tabular}{|c||c|c|c|c|}
\hline
\multicolumn{1}{|c}{\textbf{Client \hfill Server}} & 
\multicolumn{1}{c}{\textbf{$S \rightarrow C$ (\%)}} & 
\multicolumn{1}{c}{\textbf{None (\%)}} & 
\multicolumn{1}{c}{\textbf{$C \rightarrow S$ (\%)}} & 
\multicolumn{1}{c|}{\textbf{Error (\%)}} \\
\hline
\hline
CN \hfill Tor$-$Relay & 116,460 (81.52)  & 555 (0.39) & 786 (0.55) & 25,061 (17.54)\\
CN \hfill Tor$-$Dir & 8,922 (64.91)  & 31 (0.23) & 2,696 (19.61) & 2,097 (15.25)\\
CN \hfill Web & 306 (1.23)  & 15,663 (62.95) & 2,688 (10.80) & 6,226 (25.02)\\
EU \hfill Tor$-$Relay & 18 (0.20)  & 8,589 (96.79) & 22 (0.25) & 245 (2.76)\\
EU \hfill Tor$-$Dir & 2 (0.25)  & 776 (96.76) & 0 (0.00) & 24 (2.99)\\
EU \hfill Web & 19 (1.23)  & 1,333 (86.28) & 95 (6.15) & 98 (6.34)\\
NA \hfill Tor$-$Relay & 45 (0.39)  & 11,022 (94.48) & 33 (0.28) & 566 (4.85)\\
NA \hfill Tor$-$Dir & 4 (0.37)  & 1,025 (94.73) & 3 (0.28) & 50 (4.62)\\
NA \hfill Web & 32 (1.52)  & 1,794 (85.06) & 98 (4.65) & 185 (8.77)\\
\hline
\end{tabular}
\end{table*}

\subsection{Temporal and spatial association} \label{sec:mueen}

\begin{figure}[t]
\centering
\includegraphics[width=0.4\textwidth]{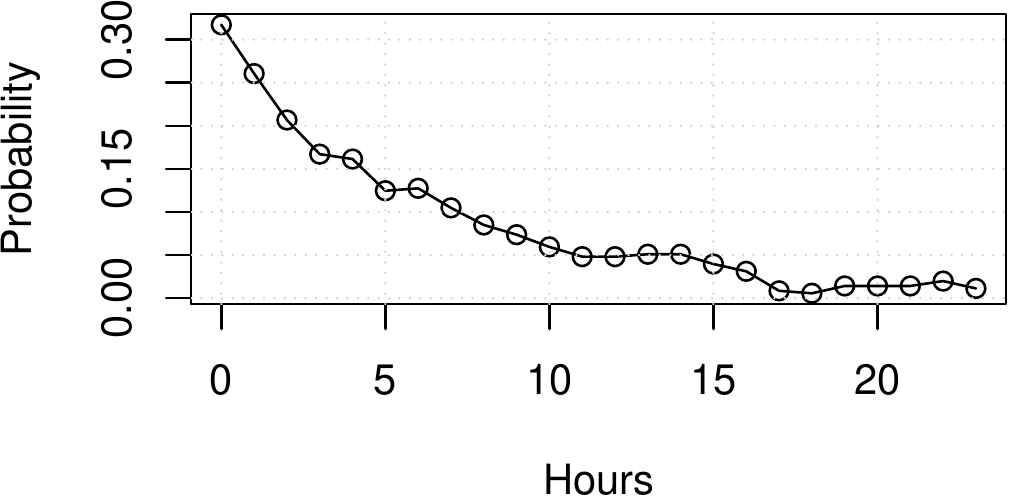}
\caption{The temporal association between cases of {\bf No-packets-dropped}.
The $x$ axis shows the amount of hours since the last {\bf No-packets-dropped}
case whereas the $y$ axis shows the probability of observing another case of
{\bf No-packets-dropped}.}
\label{fig:probability}
\end{figure}

We now seek to answer the question of whether there are any temporal or spatial
associations among the {\bf No-packets-dropped} cases observed for Tor
relays tested from within China.

Temporal association is shown in Figure~\ref{fig:probability}. The probabilities
are computed by a simple counting technique. We have the hourly count of the
number of {\bf No-packets-dropped} cases for each source. For each
occurrence of {\bf No-packets-dropped}, we check if there are other {\bf
No-packets-dropped} cases in the subsequent hours.  We use 151 sources for
this calculation, excluding the educational sources, which contained 353 {\bf
No-packets-dropped} cases in total. The final probabilities are averaged
over all sources. With the increase in the lag amount in the $x$-axis, the
probability decreases.  This shows that {\bf
No-packets-dropped} cases generally happen in bursts of hours.

\begin{figure}[t]
\centering
\includegraphics[width=0.4\textwidth]{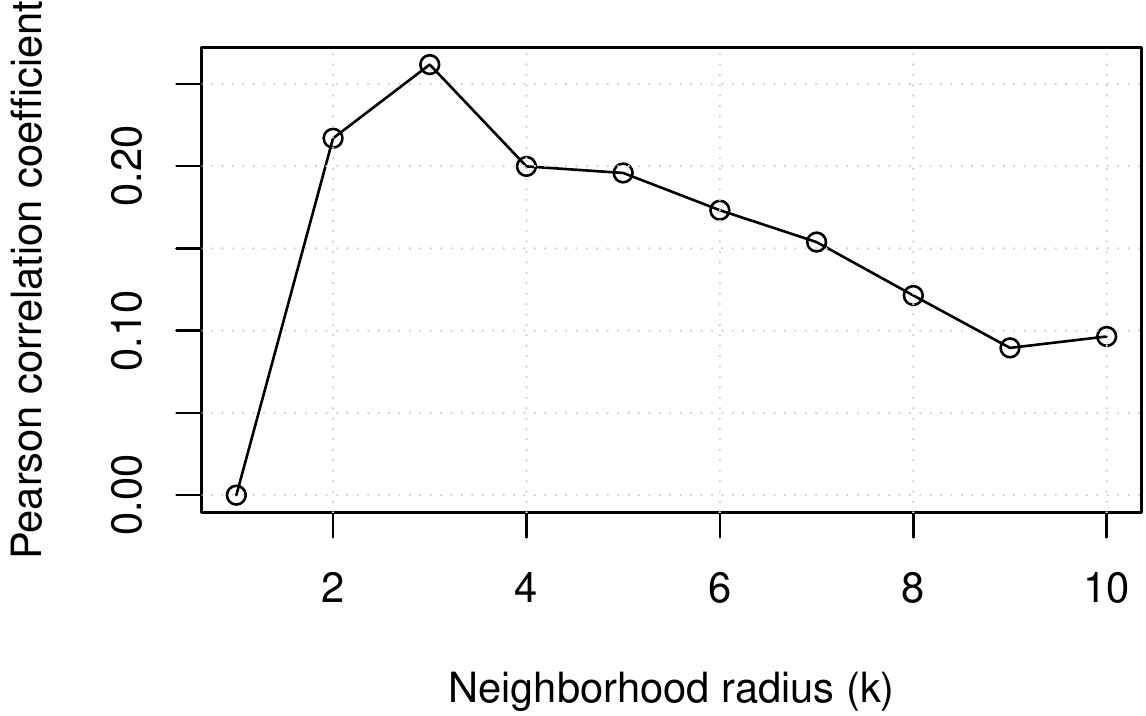}
\caption{Spatial association between clients in China. The $x$ axis shows the
neighborhood radius (k) and the $y$ axis shows the Pearson correlation
coefficient.}
\label{fig:neighborhood}
\end{figure}

Spatial association is shown in Figure~\ref{fig:neighborhood}. We use the
latitude and longitude of the sources as two-dimensional coordinates.  The
curvature of the earth is ignored while computing the distance between sources.
For every source, we find the geographically K-nearest neighboring sources and
average their count. We compute the Pearson's correlation coefficient between
the count of {\bf No-packets-dropped} cases for a source and the average
of the same for the neighboring sources. Note that Pearson's correlation has a
range of $-1.0$ to $1.0$. Our maximum observed correlation value of $0.26$ is,
therefore, a very weak positive correlation and supports the fact that there is
no significant geographical association between sources and their neighbors.
With the increase of the neighborhood radius, the correlation decreases to
below $0.1$.  Together with the fact that the cases of {\bf
No-packets-dropped} are distributed fairly evenly in all geographic
regions (see Figure ~\ref{fig:Case2Heat}), this is strong support for
Hypothesis~\ref{hypo:allover}.

\subsection{SYN backlog scans}
We began our backlog scans on 24 March 2014 and ran them twice a day with
approximately 12 hours in between the scans until 10 April 2014.  We gathered a
total of 2,094 scans and after pruning, this effort yielded 1,320 scans (63\%).


\subsubsection{Reachable Tor relays}
Out of all 1,320 backlog scans, 33 scans (2.5\%) to 12 unique IP addresses
contained the respective Tor relay's  SYN/ACK segments, indicating that no
filtering was happening.  Interestingly, 19 of these 33 scans targeted the
directory authority 128.31.0.39 on port 9131.  Only the RST scan and not the
SYN scan yielded SYN/ACKs from the directory authority.

\begin{table}
\small
\caption{Backlog scan results\label{tab:markofthebeast}.}
\centering
\begin{tabular}{|c||c|c|}
\hline
\multicolumn{1}{|c}{} &
\multicolumn{1}{c}{\textbf{RST passes}} & 
\multicolumn{1}{c|}{\textbf{RST dropped}} \\
\hline
\hline
\textbf{SYN passes} & 666 (80\%) & 39 (4.7\%) \\
\textbf{SYN dropped} & 68 (8.2\%) & 53 (6.4\%) \\
\hline
\end{tabular}
\end{table}

The results in Table~\ref{tab:markofthebeast} show that, in general, if a RST
packet passes through the GFW then a SYN packet also will.  This confirms one of
the basic assumptions behind the hybrid idle scan, and confirms
Hypothesis~\ref{hypo:synrst}.  Also, the fact that most SYNs were allowed to
pass through the GFW confirms Hypothesis~\ref{hypo:synack}.

\subsection{Traceroutes}
\label{sec:traceroute_analysis}

\begin{table*}
\small
\caption{The results of our traceroute measurements.}
\label{tab:tracer}
\centering
\begin{tabular}{|l||c|c|c|c|}
\hline
\multicolumn{1}{|c}{} &
\multicolumn{1}{c}{\textbf{EDU Randport}} &
\multicolumn{1}{c}{\textbf{EDU Torport}} &
\multicolumn{1}{c}{\textbf{COM Randport}} &
\multicolumn{1}{c|}{\textbf{COM Torport}} \\
\hline
\hline
\textbf{Stalled} & 1,061 & 1,045 & 111,133 & 163,095 \\
\textbf{Finished} & 428 & 433 & 53,479 & 429 \\
\hline
\end{tabular}
\end{table*}

Table~\ref{tab:tracer} shows the results of our traceroute measurements.  In
the table, ``EDU'' indicates that the first hop in China in the traceroute is
the educational and research network backbone, CERNET (210.250.0.0/16 or
101.4.112.0/24) or another scientific network called CSTNET (159.226.0.0/16).
``COM'' indicates that the first hop in China was a commercial backbone, one
of: CNCGROUP (219.158.0.0/16), China Telecom/CHINANET (202.97.0.0/16), China
Mobile Communications Corporation (211.136.1.0/24 or 221.176.23.0/24), or the
China Telecom Next Carrying Network backbone (50.43.0.0/16).  All other entry
points were thrown out because they were actually in Hong Kong or Pasadena, and
that usually indicated that the destination IP address was not in China or
non-Chinese routing hops had not been properly culled.  ``Tor'' means that the
source port of the SYN/ACKs sent in the traceroute was the Tor port, and
``rand'' means that the source port was another port that the GFW does not
filter.  Thus, ``Tor'' traceroutes should always stop before the destination
host if the filtering is effective on that route, and ``rand'' should reach the
destination unless there are other types of filtering in play, such as ICMP
filtering or firewalls not related to censorship.  The elements in the table
are the number of times that a traceroute reached all the way to the
destination.

Surprisingly, the educational and research networks, in particular CERNET, do
not seem to be implementing this type of filtering at all.  The ``Tor'' and
``rand'' columns are nearly identical for the ``EDU'' traceroutes.  The ``COM''
traceroutes, however, show that commercial networks are clearly censoring Tor
by dropping SYN/ACKs.  The ``rand'' traceroutes reached their destination
53,479 times, while the ``Tor'' traceroutes aimed at the same destinations only
reached the destination end host 429 times.  Similar to the hybrid idle scan
results, these failures were all over the country and for any destination IP
address where at least one failure was observed, the number of failures ranged
from 1 to 48 (\emph{i.e.}, all 48 hours of measurements).  The number of
failures in the most prominent destinations where the traceroute entered China
on a commercial background included one instance where 48 failures were
observed and two where 47 were observed.  This means that sometimes the
failures are relatively persistent, confirming
Hypothesis~\ref{hypo:persistent}. 

\begin{figure}[t]
\centering
\includegraphics[width=0.47\textwidth]{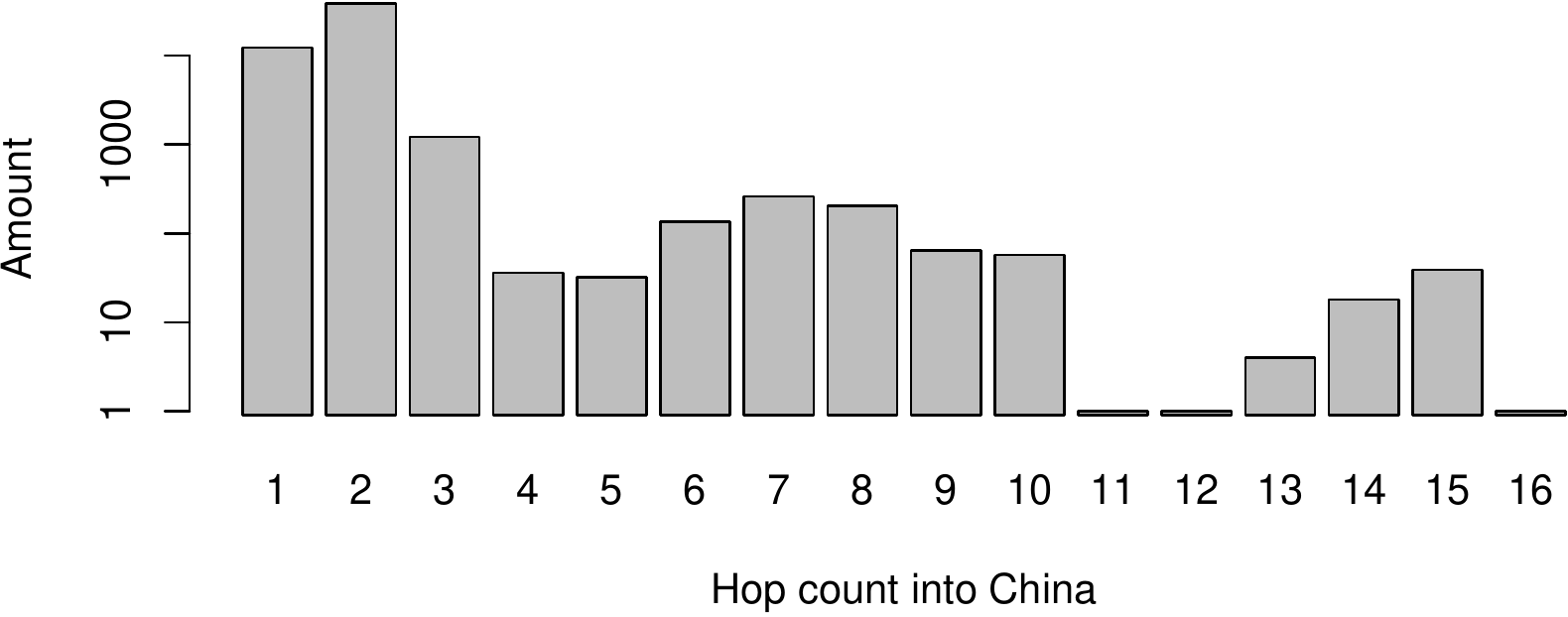}
\caption{The amount of hops (log scale) in China, our filtered traceroutes
could traverse.  For example, a hop count of five means that a traceroute could
successfully reach the fifth router inside China.}
\label{fig:hopcount}
\end{figure}

Figure~\ref{fig:hopcount} shows the amount of hops into China, filtered ``Tor''
port traceroutes traversed before stalling.  For each measurement of each
hour of each day, we only add the data to Figure~\ref{fig:hopcount} if the
``rand'' traceroute reached the destination and the ``Tor'' traceroute did not.
In most cases, the filtered packets make it two hops into China, confirming
Hypothesis~\ref{hypo:coupleofhops}.

\begin{figure}[t]
\centering
\includegraphics[width=0.47\textwidth]{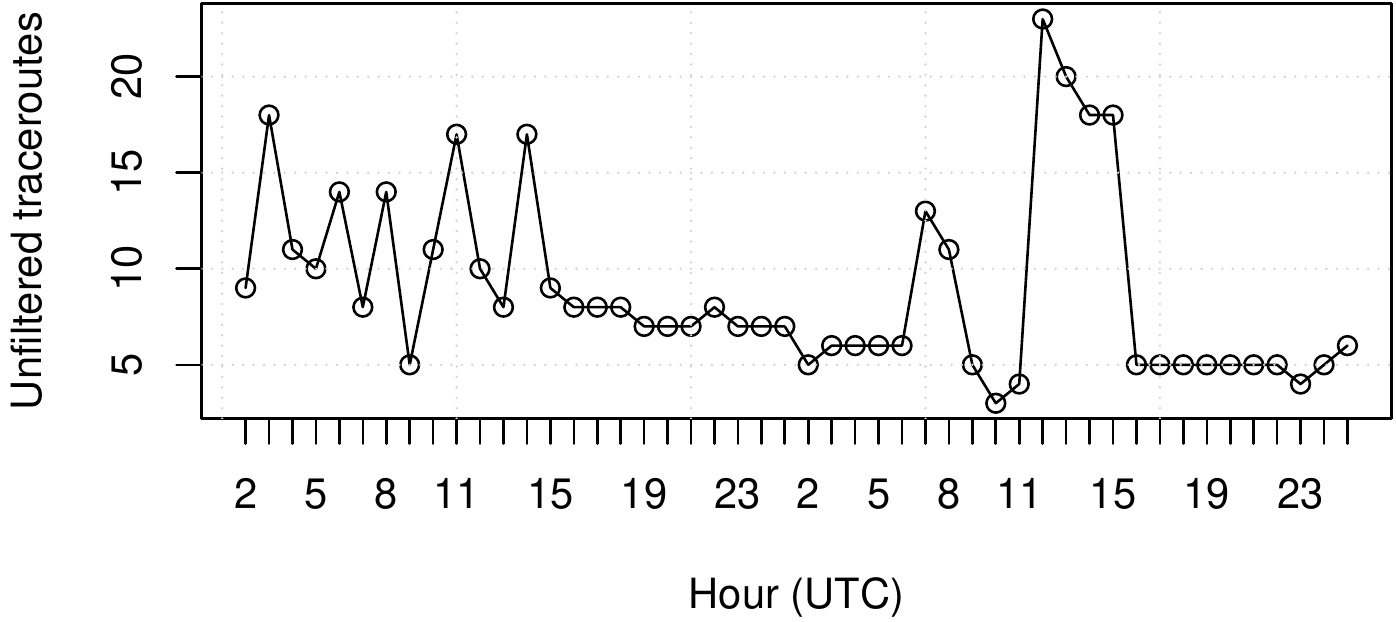}
\caption{The amount of unfiltered traceroutes from our Tor relay to clients in
China over time. A diurnal pattern is visible.}
\label{fig:diurnal}
\end{figure}

Figure~\ref{fig:diurnal} shows the number of failures for traceroutes that
entered China on the commercial network backbone, per hour.  The diurnal
patterns apparent in the figure confirm Hypothesis~\ref{hypo:diurnal}.  Note
that 02:00 UTC is 10:00 (or, 10:00 am) in Beijing.

\section{Discussion}\label{sec:discussion}
We discuss three different aspects of our work in this section: what we learned
about the filtering of Tor in China, what we learned about the architecture of
the GFW, and ethical considerations.

\subsection{Filtering of Tor in China}

\begin{figure}[t]
\centering
\includegraphics[width=0.47\textwidth]{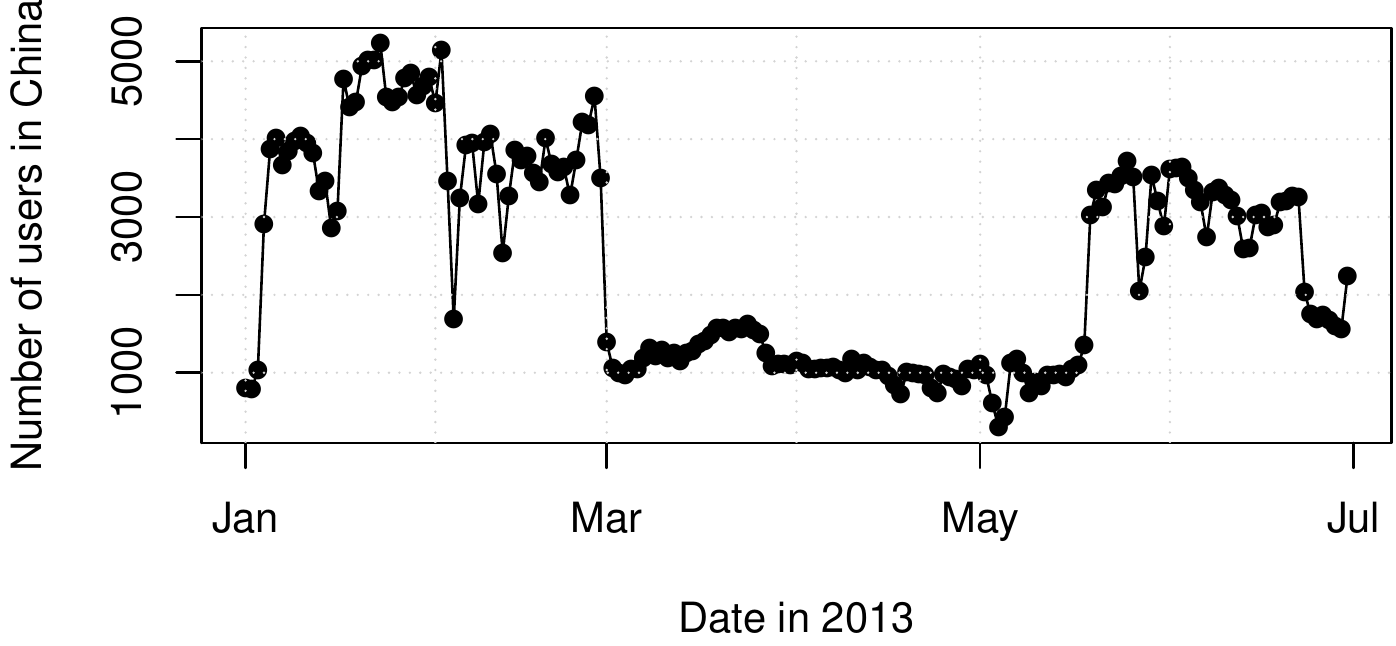}
\caption{The amount of directly connecting Tor users over the first seven
months of 2013.  The diagram shows several spikes and a ``valley'' in between
March and May.}
\label{fig:china_outages}
\end{figure}

Our results suggest that the filtering of Tor in China has several interesting
aspects, some of which may even be useful for circumvention efforts.  We showed
that the failures in the filtering occur in every part of the country, and they
are sometimes \emph{intermittent} and sometimes \emph{persistent}.  A
historical example of intermittent failures is illustrated in
Figure~\ref{fig:china_outages}.  The diagram shows the amount of directly
connecting Tor users in China in the first seven months of 2013.  A relatively
stable ``valley'' in between March an May is clearly visible.  This valley is
surrounded by significantly higher usage numbers.

We also showed that this type of filtering does not occur on CERNET, the
educational and research backbone of China's Internet.  This might suggest that
CERNET users can reach the Tor network, or it might suggest that CERNET employs
a more sophisticated method for detecting and interfering with connections to
the Tor network, perhaps something stateful and based on deep packet
inspection.

Our results raise additional questions such as ``is it possible to run a Tor
relay in China?''.  In general, the Tor network represents a complete graph.
As a result, every relay should be able to connect (and generally maintain
connections) to all other relays in the network.  Furthermore, relays must be
able to connect to the directory authorities in order to upload their server
descriptors.  If CERNET is indeed whitelisted, a Tor relay inside CERNET might
be able to successfully join the Tor network.  In addition, previous research
suggested that domestic Tor traffic in China is not subject to
blocking~\cite{Winter2012}.  If filtering indeed happens at the Internet
exchange point (IXP) level, as suggested by our data, it is not surprising that
the GFW is generally unable to filter domestic network traffic as it typically
does not reach IXP level\footnote{We ignore routing phenomena such as
``boomerang routing''.} and is of significantly higher volume than
international traffic.  As a result, functioning Tor relays or bridges inside
CERNET might be able to connect users in China to the rest of the Tor network.

\subsection{The architecture of the GFW}

Our results also shed light on the architecture of the GFW, at least with
respect to the mechanism that blacklists IP address/TCP port pairs.  As
discussed in Section~\ref{sec:background}, the three theories about how the GFW
is architected are that \emph{1)} the filtering occurs at choke points where
undersea cables enter the country, \emph{2)} the filtering occurs in the
backbone in large IXPs, and \emph{3)} the filtering occurs at a regional level.
While our results show some filtering occurring many hops into China and some
filtering occurring before packets can even enter China, the majority of the
filtering happens about two hops into China (presumably at the large IXP in
Beijing).  Thus, Hypothesis~\ref{hypo:coupleofhops} is most consistent with the
theory that the filtering occurs in the backbone.  Note that this observation is
in accordance with other recent research efforts which focused on the GFW's DNS
injection~\cite{Anonymous2014}.  The small amount of routes that are filtered at
the provincial level, which were also observed by Xu \emph{et
al.}~\cite{Xu2011}, can be explained by the strategy
employed by China's formerly second-largest ISP, CNCGROUP, which was recently
bought by the largest (CHINANET).

While whitelisting would appear as persistent failures in the filtering and the
filtering apparatus getting overloaded with traffic would appear as intermittent
failures, the mix of intermittent failures and diurnal patterns with persistent
failures suggests that routing is a major reason why the filtering fails.
Hypotheses~\ref{hypo:persistent} and~\ref{hypo:diurnal} are most consistent with
the theory that the filtering occurs in the backbone, because provincial
networks in China are very hierarchical~\cite{10.1109/TPDS.2012.271} and
undersea cables are few in number~\cite{cablemap}.
Hypothesis~\ref{hypo:allover} is also most consistent with backbone-level
filtering for this reason.

\subsection{Ethical considerations}
\label{sec:ethics}
Our work has two ethical considerations that need to be discussed.  First, our
SYN backlog scans briefly fill a Tor relay's backlog in order to be able to
observe packet drops.  A full backlog can prevent a relay from accepting new
TCP connections or cause the use of SYN cookies which can lead to reduced
throughput.  To prevent relays from using SYN cookies, we adapted our scan
parameters to minimize the risk of completely filling a relay's SYN backlog.
SYN cookies typically do not support scaled flow control windows, which is why
we made every effort to avoid them.  In general, the rate at which we are
sending SYN packets, without intention of completing a connection, is not
enough to create a denial-of-service condition on any modern network stack.
For an interesting discussion about ethical issues related to port scans in
general, we refer the reader to Durumeric \emph{et al.}~\cite{zmap13}.

Second, our idle scans create unsolicited traffic between a client and a
server.  This traffic---which can be observed by the censor---is only
SYN/ACKs from the server to the client and RSTs from the client to the server.
As a result, we are not causing any meaningful communication other than
background noise as it is also caused by port scanning activity.  While one may
conceptualize the hybrid idle scan technique as providing the ability to
conscript a client into performing tests for us, in reality the traffic between
the server and the client is no different from if the server chose to send
SYN/ACKs to the client.  Thus, in terms of the traffic that the censor sees,
the hybrid idle scan technique is no different from if Tor relay operators
performed simple connectivity measurements by directly sending SYN/ACKs.

\section{Related Work}\label{sec:relatedwork}
As our work employs network inference techniques in order to measure the
reachability of the Tor anonymity network, we divide related work in two
subsections.  The first subsection focuses on similar network inference
techniques, and the second discusses the Great Firewall of China and Internet
censorship measurements in general. 

\subsection{Network inference techniques}

There has been a fair amount of work on utilizing side channels in TCP/IP
network stacks.  Antirez's seminal IPID idle scan from
1998~\cite{nmapbook,antirez} and other work on idle scans~\cite{roya} focus on
network security.  Qian \emph{et al.}~\cite{Qian:2012:OTS:2310656.2310690} show
that some firewalls exhibit behavior that can be used to infer sequence numbers
and hijack connections.  Chen \emph{et
al.}~\cite{Chen:2005:EIF:2150193.2150205} use the IPID field to perform
advanced inferences, such as the amount of internal traffic generated by a
server, the number of servers in a load-balanced setting, and one-way delays.
Morbitzer~\cite{morbitzerthesis} explores idle scans in IPv6.
Queen~\cite{Wang:2009:QEP:1532940.1532949} utilizes recursive DNS queries to
estimate the packet loss between a pair of arbitrary hosts by measuring the
packet loss between their respective DNS servers.  Reverse
traceroute~\cite{Katz-Bassett:2010:RT:1855711.1855726} is an interesting
application of indirect methods for Internet measurement.  

Passively identifying hosts that have no routable IP address and are hidden by
network address translation~\cite{natted,devfingerprint} is a related problem to
inferring connectivity of hosts.

iPlane~\cite{Madhyastha:2006:IIP:1298455.1298490} sends packets from PlanetLab
nodes to carefully chosen hosts, and then compounds loss on specific routes to
estimate the packet loss between arbitrary endpoints.  The view of the network
is fundamentally limited to the perspective of the measurement machine, however.
Queen~\cite{Wang:2009:QEP:1532940.1532949} utilizes recursive DNS queries to
measure the packet loss between a pair of DNS servers, and extrapolates from
this to estimate the packet loss rate between arbitrary hosts.  

To the best of our knowledge, our work is the first use of idle scan inference
techniques for a large-scale Internet measurement study where the data
collected gives a view of the network from the perspective of a very large
number of clients distributed over a large country.  Platforms such as
DIMES~\cite{dimesmap}, M-Lab~\cite{mlabmap}, PlanetLab~\cite{planetlabmap}, and
RIPE Atlas~\cite{RIPEmap,Anderson2014} have traditionally been the only way to
measure from the perspective of a large number of clients, but they can be very
limited, especially in non-Western regions of the Internet such as China.  Our
work overcomes a fundamental limitation of Internet measurement: that
measurements traditionally have only been possible from the perspective of the
measurement machines controlled directly by researchers.

\subsection{The Great Firewall of China}

The Great Firewall of China was first described in an article in 2600
magazine~\cite{gfw2600}.  In 2006, Clayton, Murdoch, and Watson investigated the
firewall's keyword filtering mechanism and demonstrated that it can by
circumvented by simply ignoring the firewall's injected RST
segments~\cite{Clayton2006}.  Clayton \emph{et al.}'s study was limited to how
the filtering works.  \emph{What} it filters was covered by Crandall \emph{et
al.} in 2007~\cite{conceptdoppler}, along with more details about routing.
Using latent semantic analysis, the authors bootstrapped a set of 122 keywords
which were used to probe the firewall over time.  The study also shows that
filtering is probably not happening at the border of China's Internet.  Xu, Mao,
and Halderman made an effort to pinpoint where exactly the filtering is
happening~\cite{Xu2011}.  The authors came to the conclusion that most filtering
is happening in border ASes but some filtering is also happening in provincial
networks.  Park and Crandall revisited the GFW's keyword filtering mechanism and
discussed why the filtering of HTML responses was discontinued in late
2008~\cite{Park2010}.

In addition to topology and HTTP filtering, another direction of research
focused on how the GFW operates on the TCP/IP layer.  In 2006, Clayton \emph{et
al.} already showed that the GFW is terminating suspicious HTTP requests using
injected RST segments.  Weaver, Sommer, and Paxson showed that it is possible to
not only distinguish genuine from injected RST segments but also to fingerprint
networking devices injecting the segments~\cite{Weaver2009}.  More recently in
2013, Khattak \emph{et al.} probed the GFW in order to find evasion
opportunities on the TCP/IP layer~\cite{Khattak2013}.  Resorting to techniques
first discussed by Ptacek and Newsham in 1998~\cite{ptaceknewsham}, the authors
showed that there are numerous evasion opportunities when crafting TCP and IP
packets.  Similarly, Winter and Lindskog showed in 2012 that packet
fragmentation used to be sufficient to evade the GFW's deep packet
inspection~\cite{Winter2012}.

In addition to the design and topology of the GFW, some work focused on how the
GFW blocks application protocols other than HTTP.  In 2007, Lowe, Winters, and
Marcus showed that the GFW is also conducting DNS poisoning~\cite{Lowe2007}.  A
more comprehensive study was conducted by anonymous authors in
2012~\cite{Anonymous2012} and 2014~\cite{Anonymous2014}.  The authors sent DNS
queries to several million IP addresses in China, thereby demonstrating that
the GFW's DNS poisoning causes collateral damage, \emph{i.e.}, interferes with
communication outside China.  A follow-up study was conducted in 2014---also by
anonymous authors~\cite{Anonymous2014}.  The authors probed a large body of
domain names to determine how filtering changes over time.  Furthermore, the
authors approximated the location of DNS injectors.  Interestingly, their
results are similar to ours and they write that ``In most cases, the injecting
interface manifested at either 2 (18.3\%) or 3 (54.6\%) hops inside China''
(cf.~\ref{sec:traceroute_analysis}).

Most work discussed so far treated the
firewall as a monolithic entity.  Wright showed in 2012 that there are regional
variations in DNS poisoning, thus suggesting that censorship should be
investigated on a more fine-grained level with attention to geographical
diversity in measurements~\cite{Wright2012}.  In addition to DNS and HTTP, the
GFW is known to block the Tor anonymity network.  Using a VPS in China, Winter
and Lindskog~\cite{Winter2012} investigated how the firewall's active probing
infrastructure is used to dynamically block Tor bridges.

In terms of Internet censorship measurements not aimed at the GFW, there is a
growing body of work but two works in particular are notable from an Internet
measurement perspective.  Dainotti \emph{et
al.}~\cite{Dainotti:2011:ACI:2068816.2068818} analyze several Internet
disruption events that were censorship-related using various data sources from
both the control and data planes.  Dalek \emph{et al.}~\cite{imc2013a} present a
method for identifying externally visible evidence of URL filtering.

The most notable difference to previous work is that our measurement techniques
do not require control over either machine which is part of censored
communication.  While that enables large-scale distributed studies, it comes at
the cost of reduced flexibility.

\section{Conclusion}\label{sec:conclusion}
In this paper, we have characterized the mechanism that the Great Firewall of
China uses to block the Tor network using a hybrid idle scan that can measure
connectivity from the perspective of many clients all over China.  We have also
presented a novel SYN backlog idle scan that can infer packets received by a
server without causing denial of service.  These novel Internet measurement
techniques open up whole new possibilities in terms of being able to measure
the Internet from the perspective of arbitrary clients and servers.  This is
extremely important when it comes to characterizing and documenting Internet
censorship around the world, because of the difficulty in finding volunteers
geographically dispersed throughout a country.

We also evaluated our techniques which led to several new insights about the
inner workings of the Great Firewall.  Our data shows that \emph{1)} at least
several machines inside CERNET (China Education and Research Network) are able
to connect to Tor relays, \emph{2)} filtering seems to be centralized at the
IXP level, and \emph{3)} filtering is quite reliable with the Tor network being
either almost completely reachable or almost completely blocked in different
parts of the country.

Our code is available at: \url{http://cs.unm.edu/~royaen/gfw/}.


\section*{Acknowledgments}
We would like to thank Kasra Manavi for his valuable feedback. This material is based upon work supported
by the National Science Foundation under Grant Nos. \#0844880, \#1017602,
\#0905177, and \#1314297.  The author from Karlstad University was supported by
a research grant from Internetfonden.

\bibliographystyle{IEEEtran}
\bibliography{references}

\end{document}